\documentclass[twocolumn,superscriptaddress,floatfix,showpacs]{revtex4-1}
\usepackage{graphics,amssymb,amsmath,epsfig,color}
\usepackage{graphicx}

\newcommand{\be}{\begin{equation}} \newcommand{\ee}{\end{equation}}
\newcommand{\bea}{\begin{eqnarray}} \newcommand{\eea}{\end{eqnarray}}

\begin{document}

\title{Agglomerative Percolation on Bipartite Networks: A Novel Type of Spontaneous Symmetry Breaking}
\author{Hon Wai Lau} \affiliation{Complexity Science Group, University of Calgary, Calgary T2N 1N4, Canada}
\author{Maya Paczuski} \affiliation{Complexity Science Group, University of Calgary, Calgary T2N 1N4, Canada}
\author{Peter Grassberger} \affiliation{Complexity Science Group, University of Calgary, Calgary T2N 1N4, Canada}
\date{\today}

\begin{abstract}

Ordinary bond percolation (OP) can be viewed as a process where clusters grow by joining them 
pairwise, by adding links chosen randomly one by one from a set of predefined `virtual' links. 
In contrast, in agglomerative percolation (AP) clusters grow by choosing randomly a `target 
cluster' and joining it with {\it all} its neighbors, as defined by the same set of virtual links.
Previous studies showed that AP is in different universality classes from OP for several types 
of (virtual) networks (linear chains, trees, Erd\"os-R\'enyi networks), but most surprising were 
the results for 2-d lattices: While AP on the triangular lattice was found to be in the OP 
universality class, it behaved completely differently on the square lattice. In the present
paper we explain this striking violation of universality by invoking bipartivity. While 
the square lattice is a bipartite graph, the triangular lattice is not. In conformity with 
this we show that AP on the honeycomb and simple cubic (3-d) lattices -- both of which are 
bipartite -- are also not in the OP universality classes. More precisely, we claim that 
this violation of universality is basically due to a $Z_2$ symmetry that is spontaneously 
broken at the percolation threshold. We also discuss AP on bipartite random networks 
and suitable generalizations of AP on $k-$partite graphs.

\end{abstract}

\maketitle

\section{Introduction}

Percolation was until recently considered a mature subject that held few surprises, but 
this has changed dramatically during the last few years. Recent discoveries that widened
enormously the scope of different behaviors at the percolation threshold include 
infinite order transitions in growing networks \cite{callaway}, supposedly first order 
transitions in Achlioptas processes \cite{Achli-2009} (that are actually 
continuous \cite{Costa-2010,Riordan} but show very unusual finite size 
behavior \cite{Grassberger}), and real first order transitions in interdependent networks
\cite{Buldy-2010,Parshani-2010,Son-2011,Son-PRL-2011}. Another class of ``non-classical" percolation
models, inspired by attempts to formulate a renormalization group for networks \cite{Song-2005,Radic-2009}, 
was introduced in \cite{Bizh-2010,Christen-2010,Son-2010,Son-EPL-2011,Bizhani-PRE-2011} 
and is called `agglomerative percolation' (AP). 

The prototype model in the ordinary percolation (OP) universality class is bond percolation
\cite{Stau-1994}.
There one starts with a set of $N$ nodes and a set of `virtual' links between them, i.e. links 
that {\it can} be placed but that are not yet put down. One then performs a process where 
one repeatedly picks at random one of the virtual bonds and realizes it, i.e. actually 
links the two nodes. A giant cluster appears with probability one in the limit $N\to\infty$, 
when the density $p=M/N$ of links ($M$ is here the number of realized links) exceeds a 
threshold $p_c$ whose value depends on the topology of the network. The behavior at $p\approx p_c$
is governed by `universal' scaling laws, i.e. by scaling laws with exponents that depend only on 
few gross properties of the network. A typical example is that the 
universality class of OP on regular $d-$dimensional lattices depends on $d$ but not on
the lattice type. For example, OP on triangular and square lattices (both have $d=2$) are in
the same universality class.

AP differs from OP in that clusters do not grow by establishing links one by one. Rather,
one picks a `target' cluster at random (irrespective of its mass; we are dealing here with 
model (a) in the classification of \cite{Christen-2010}) and joins it with {\it all} its 
neighbors, where neighborhoods are defined by the virtual links. The new combined cluster
is then linked to all neighbors of its constituents. AP can be solved rigorously on 1-d linear 
chains \cite{Son-2010,Son-EPL-2011}, where it is found to be in a different 
universality class from OP. Although a similarly complete mathematical analysis is not possible 
on random graphs, both numerics and non-rigorous analytical arguments show that the same 
is true for `critical' trees \cite{Bizh-2010} and Erd\"os-R\'enyi graphs \cite{Bizhani-PRE-2011}.

In contrast to these cases that establish AP as a novel phenomenon but do not present 
big surprises, the behavior on 2-d regular lattices \cite{Christen-2010} is extremely 
surprising: While AP on the triangular lattice is clearly in the OP universality class
(with only some minor caveats), it behaves completely different on the square lattice. 
There the average cluster size at criticality diverges as the system size $L$ increases
(it stays finite for all realizations of OP on any regular lattice), the fractal dimension
of the incipient giant cluster is $D_f=2$ ($D_f=91/48\approx 1.90$ for OP), and the cluster 
mass distribution obeys a power law with power $\tau=2$ ($\tau=187/91\approx 2.055$ for OP).
This blatant violation of universality -- one of the most cherished results of renormalization
group theory -- is, as far as we know, unprecedented.

As we said above, gross topological features of the network (such as dimensionality in case 
of regular lattices, the correlations between links induced by growing networks \cite{callaway}, 
and finite ramification in hierarchical graphs \cite{Boettcher-2011}) are one set of 
properties that determine universality classes. The other features that determine universality 
classes in general are symmetries of the order parameter: The Ising and Heisenberg models are in 
different universality classes, e.g., because the order parameter is a scalar in the 
first and a 3-d vector in the second. Could it be that the non-universality of AP
results from a similar symmetry? At first sight this seems unlikely, because the order 
parameter (the density of the giant cluster) is a scalar in any percolation model. 
Moreover, in order for a symmetry to affect the universality class it has to be broken 
spontaneously at the phase transition.

In the following we show that it is indeed the latter scenario that leads to the 
non-universality of AP on square and triangular lattices, and the symmetry that is 
spontaneously broken at the AP threshold on the square lattice is a $Z_2$ symmetry 
resulting from {\it bipartivity}. A graph is bipartite, if the set ${\cal N}$ of nodes
can be split into two disjoint subsets, ${\cal N} = {\cal N}_1 \sqcup {\cal N}_2$, 
such that all links are connecting a node in ${\cal N}_1$ with a node in ${\cal N}_2$, 
and there are no links within ${\cal N}_1$ or within ${\cal N}_2$. A square lattice 
is bipartite (as illustrated by the black/white colors of a checkerboard), but a 
triangular lattice is not. Following this example, we will in the following speak 
of the different {\it colors} of the sets ${\cal N}_1$ and ${\cal N}_2$. The initial
state of the AP process on a square lattice (where each site is a cluster) is color 
symmetric. If the AP cluster joining process is such that we can attribute a definite color
to any cluster (even when it is not a single site), then the state remains color symmetric
until we reach a state with a giant cluster. In this state the color symmetry is 
obviously broken.

In Sec. 2 we shortly review the evidence for non-universality given in 
\cite{Christen-2010}, In Sec.~3 we present new results which show that AP on square lattices 
behaves even more strange than found before. There we also present numerical 
results for the honeycomb and simple cubic lattices, both of which are bipartite 
and show similar anomalies as the square lattice. The detailed explanation why bipartivity
leads to these results is given in Sec. 4. Random bipartite networks are shortly 
treated in Sec.~5. Possible generalizations to $k$-partite graphs with $k>2$ are 
discussed in Sec.~6, while we finish with our conclusions in Sec.~7. 

\section{Agglomerative Percolation: Definition, implementation, and review of previous results}

We start with a graph with $N$ nodes and $M$ links. Clusters are defined trivially in 
this initial state, i.e. each node is its own cluster. AP is then defined by repeating 
the following step until one single cluster is left:

1) Pick randomly one of the clusters with uniform probability;\\
2) Join this `target cluster' with all its neighboring clusters, where two clusters $C_1$
and $C_2$ are `neighbors', if there exist a pair of nodes $i\in C_1$ and $j\in C_2$ that 
are joined by a link;\\

As described in \cite{Christen-2010}, this is implemented most efficiently with the 
Newman-Ziff algorithm \cite{Newman-2001} that uses pointers to point to the ``roots" of 
clusters, augmented by a breadth-first search to find all neighbors of the target.

In ordinary bond percolation one usually takes as control parameter $p$ the number of 
established (i.e. non-virtual) links, divided by the number of all possible links 
(including virtual ones). This is not practical for AP. Rather, we use as in \cite{Ziff-2009,Christen-2010} the 
number $n$ of clusters per node. It was checked carefully in these papers that
using $n$ instead of $p$ as a control parameter in ordinary bond percolation is 
perfectly legitimate, since one is a smooth monotonic (decreasing) function of the 
other.

In \cite{Christen-2010}, AP was studied on two different 2-d lattices. Helical boundary 
conditions were used for both, i.e. sites are labeled by a single index $i$ with 
$i \equiv (i \;\;{\rm mod}\;\; L^2$, where $L$ is the lattice size). For the square 
lattice the four neighbors of site $i$ are $i\pm 1$ and $i\pm L$, while there are 
two additional neighbors $i\pm (L+1)$ for the triangular lattice. This seemingly minor 
difference has dramatic consequences. While AP on the triangular lattice is (within
statistical errors, and with one minor caveat that was easily understood) in the 
universality class of OP, this is obviously not the case for the square lattice. Among 
other results, the following results were found:

\begin{figure}
\includegraphics[width=0.6\textwidth]{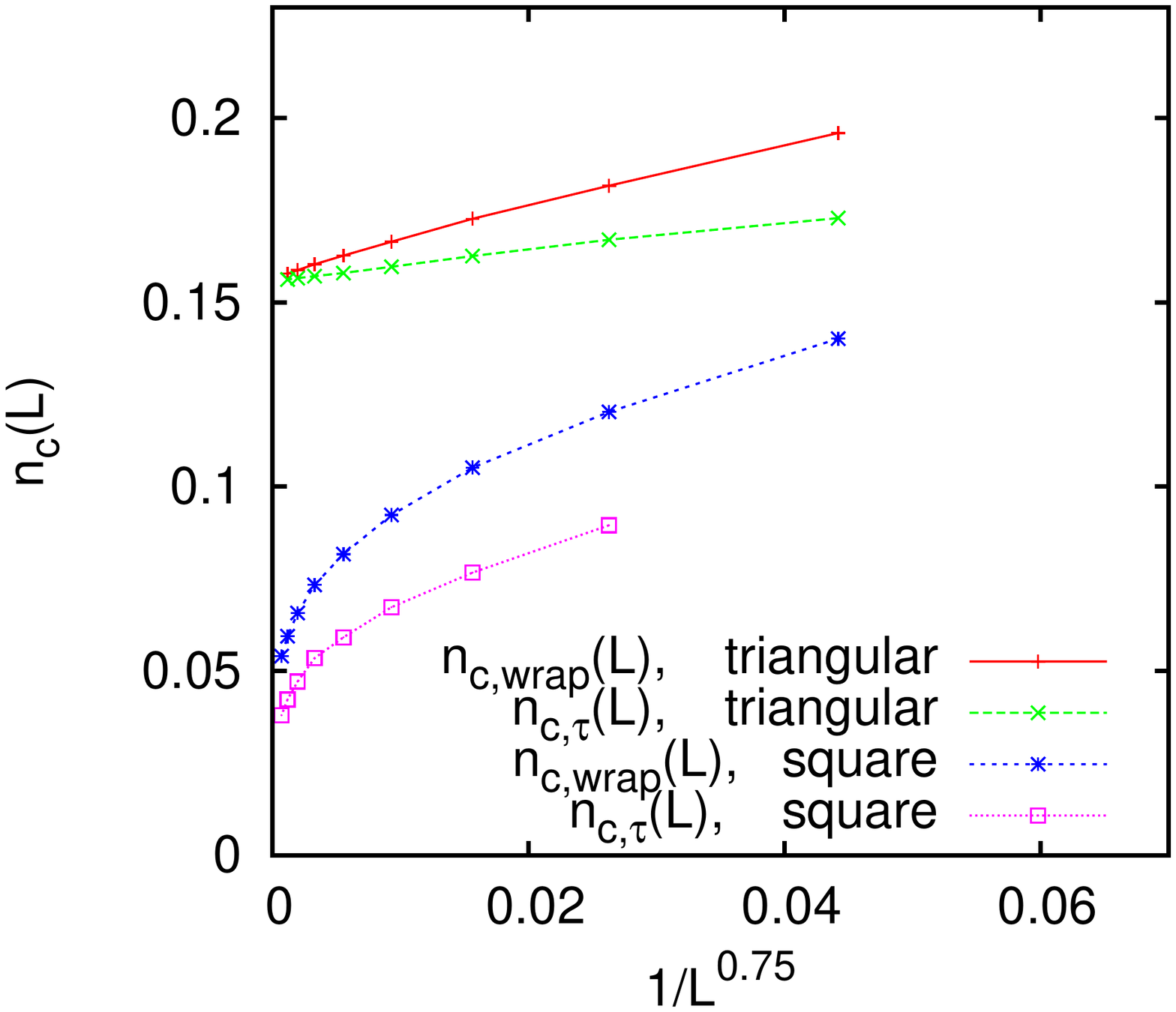}
\vglue -.5cm
\caption{(Color online) Effective critical cluster densities for AP on finite 2-d lattices versus
   $L^{-3/4}$, where $L$ is the lattice size. For ordinary percolation, where $n_c(L) -n_c
   \sim L^{-1/\nu}$ with $\nu=4/3$, this should give straight lines. For each lattice type (triangular:
   upper pair of curves; square: lower pair of curves) we show results obtained with two different
   operational definitions for the critical point: (i) Maximal range of the power law
   $P_n(m) \sim m^{-\tau}$, and (ii) the probability to have a cluster that wraps around a lattice
   with helical boundary conditions is equal to 1/2. The corresponding values of $n_c(L)$ are called
   $n_{c,\tau}(L)$ and $n_{c,{\rm wrap}}(L)$ (from Ref.~\cite{Christen-2010}).}
   \label{fig1}
\end{figure}

\begin{itemize}
\item The effective percolation threshold -- measured either by the probability that a 
cluster wraps around the lattice, or via the best scaling law 
\be 
   P(m)\sim m^{-\tau}              \label{tau}
\ee
for the probability distribution of cluster masses $m$ -- depends strongly on $L$. For 
OP this dependence is governed by the correlation length exponent $\nu$ via 
\be
   p_c - p_c(L) \propto  n_c(L) - n_c \propto L^{-1/\nu}
\ee
with $\nu = 4/3$ and $n_c >0$. The latter means, in particular, that the average cluster
size is finite at criticality. For AP on the square lattice a parameterization like this would 
give $\nu = 0$. More precisely, $n_c(L)$ seems to decrease logarithmically
to a value $n_c=0$, i.e. the average cluster at criticality
(and in the limit $L\to\infty$) is zero. This is summarized in Fig.~\ref{fig1}.

\item The exponent $\tau$ in Eq.~(\ref{tau}), which is $187/91 = 2.0549\ldots$ for OP, seems to 
be $<2$ at first sight. But it slowly increases with $L$, and it was argued 
that the exact value is $\tau=2$.

\item Similarly, the fractal dimension of the largest cluster was measured as $\approx 1.95$,
while it is $D_f= 91/48 = 1.9858 \ldots$ for OP. It also increases slowly with $L$, and 
it was conjectured that also $D_f = 2$.

\item Let $p_{\rm wrap}(n,L)$ be the probability that there exists a cluster that wraps 
along the vertical direction on 
a lattice of size $L$, when there are $nL^2$ clusters. The distribution $dp_{\rm wrap}(n) /dn$ 
is universal for OP. For AP on triangular lattices it develops a weak tail for small $n$ 
(this is the easily explained caveat mentioned above), but for AP on the square lattice it 
has a very fat tail for small $n$. Thus there is a high probability that even at very late
stages in the agglomeration process, when only few clusters remain, none of them has yet wrapped.
\end{itemize}

\section{Additional numerical results}

\subsection{Square lattice}

\subsubsection{Periodic boundary conditions}

Helical b.c. were used in \cite{Christen-2010} simply for convenience (they are more easy to
code than periodic ones), and it was assumed that the small difference with strictly periodic
b.c. should be without any consequences. This is not true. Not only is there a large
difference between helical and strictly periodic b.c.'s (even for the largest values of $L$
that we could check), but for the latter there is an even stronger difference between even and odd $L$.

\begin{figure}
\includegraphics[width=0.5\textwidth]{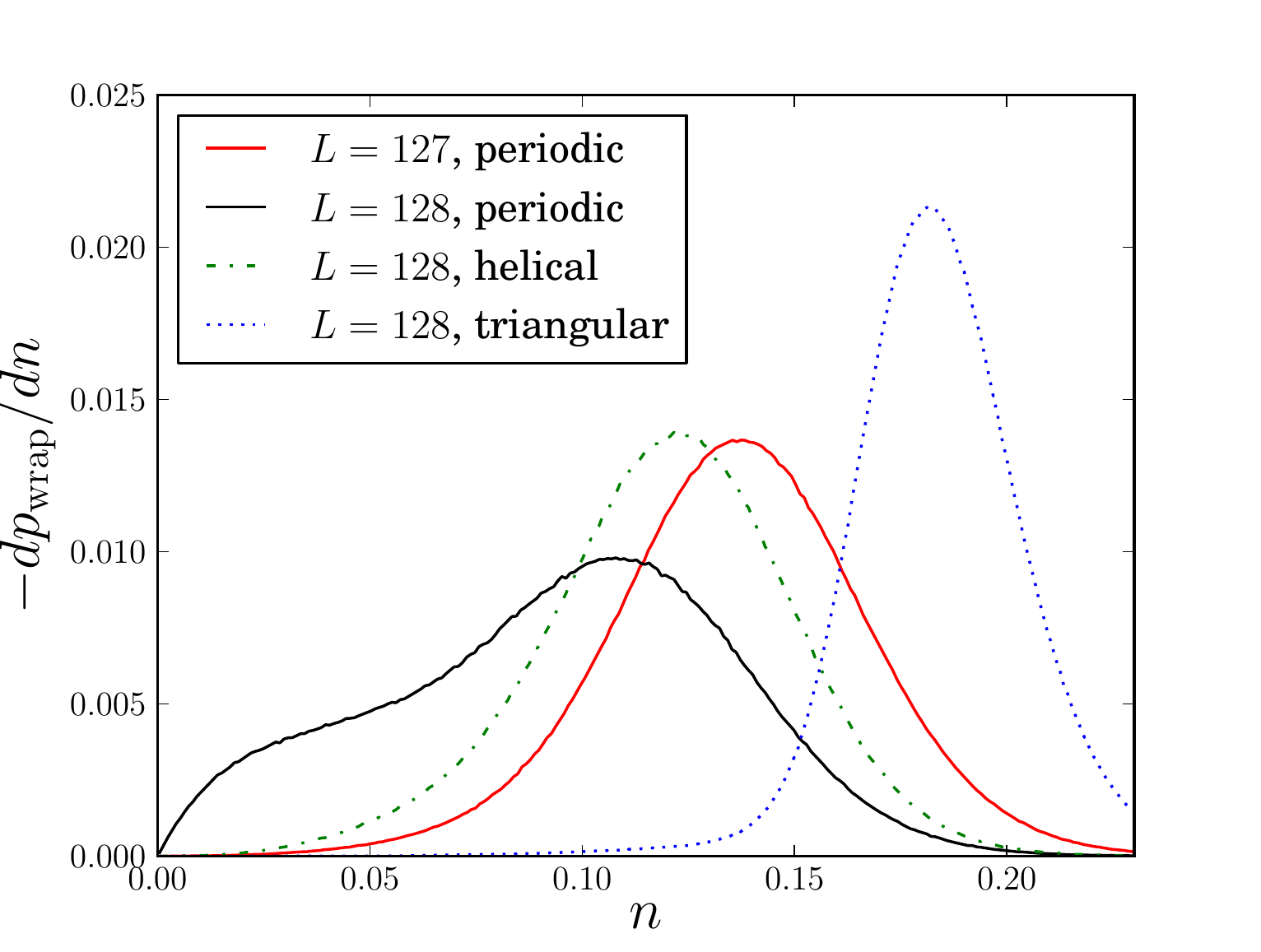}
\caption{\label{dpwrap_dn} (Color online) Wrapping probability density $dp_{\mathrm{wrap}}/dn$
for 2D square lattices with different boundary condition, and with sizes differing by just 
one unit, compared to similar results for triangular lattices. All curves for square lattices 
have peaks at smaller values than for triangular lattices and have more heavy 
left hand tails, but this is most pronounced for strictly periodic b.c. with even $L$.}
\end{figure}

\begin{figure}
\includegraphics[width=0.5\textwidth]{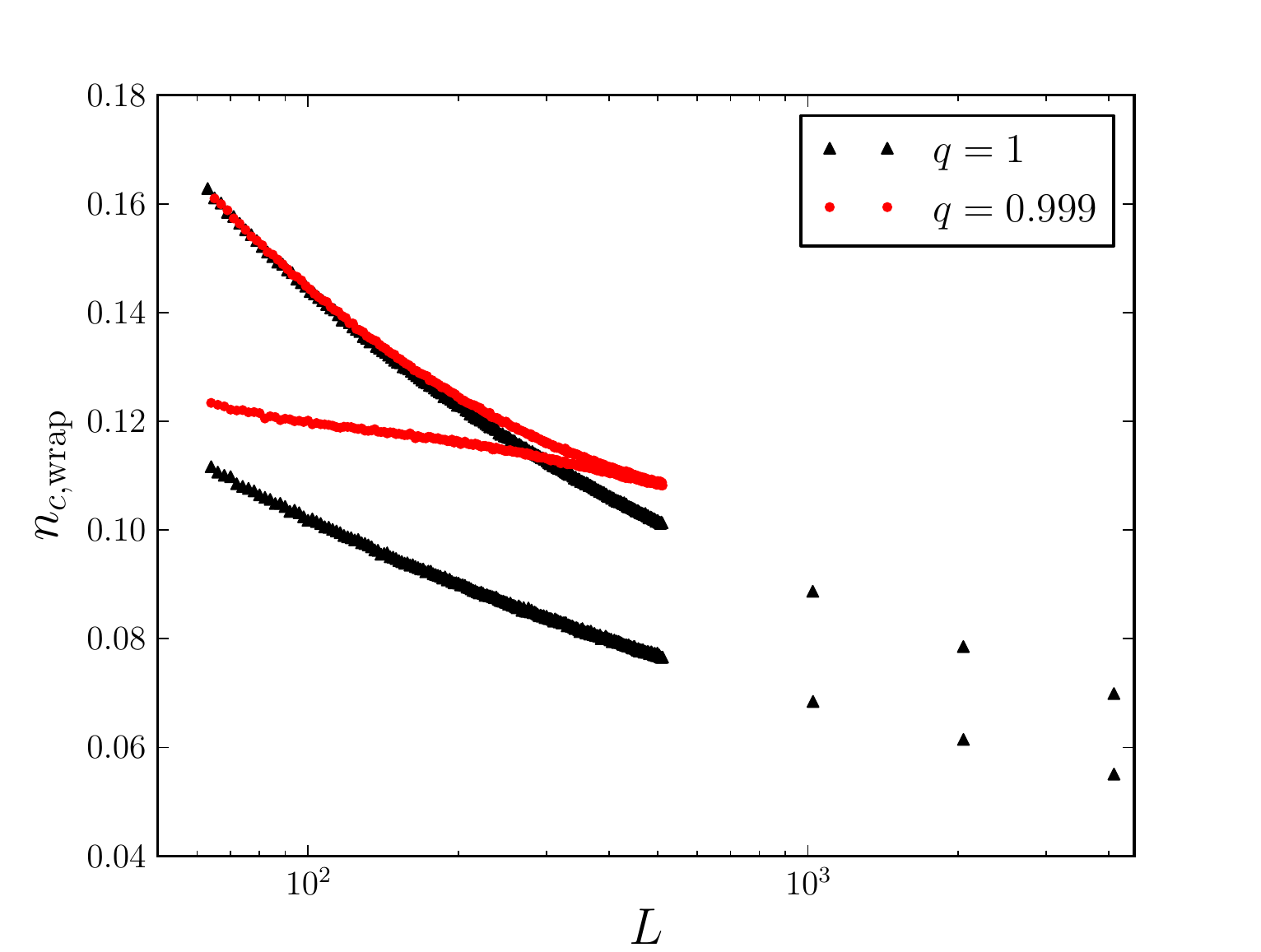}
\caption{\label{nwrap_L-2Dsq} (Color online) Black triangles: Cluster densities $n$ at which 
$p_{\rm wrap}(n)=1/2$, plotted against $L$, for square lattices with periodic b.c. 
The upper curve is for odd sizes, and the lower curve is for even sizes. Red dots: Analogous 
results for a slightly modified model, where each neighboring cluster agglomerates with the
target cluster only with probability $q<1$. In the present case $q=0.999$.}
\end{figure}

In Fig.~\ref{dpwrap_dn} we show $dp_{\rm wrap}(n) /dn$ for four different cases: \\
(i) Triangular lattices. The shape of this curve is practically indistinguishable from ordinary
percolation, and serves as a reference for the latter.\\
(ii) Square lattices of size $128\times 128$ with helical b.c. Compared to the triangular 
lattice, there is a much fatter left hand tail of the distribution, i.e. there are many more
realizations where no cluster has yet wrapped, although the number of clusters is very small.\\
(iii) Square lattices of the same size, but with periodic b.c. Now the left hand tail is even 
more fat. Indeed, for this lattice size one finds realizations with $\leq 10$ clusters, none
of which has yet wrapped.\\
(iv) Square lattices of sizes $127\times 127$ with periodic b.c. 
We see a huge difference, in spite of the small change in $L$, making the results similar to those 
for helical b.c. Obviously, this indicates a distinction between even and odd $L$. 

The last conclusion is confirmed by Fig.~\ref{nwrap_L-2Dsq}. There we show the values of $n$ where 
half of the configurations have wrapping clusters, $p_{\rm wrap}(n)=1/2$ (black triangles; the 
red dots will be discussed in subsection \ref{finite_q}). These data confirm that the difference 
between even and odd $L$ persists even to our largest systems, where the difference in $L$ between 
the two is less than 0.025 per cent.

\subsubsection{Finite agglomeration probability}
    \label{finite_q}

In AP, {\it all} neighbors of a chosen target are included in each agglomeration step. In 
contrast, bond percolation can be viewed as the limit $q\to 0$ of a model where each neighbor is 
included with probability $q$. One might then wonder where the cross-over from OP to AP happens. Is 
it for $q\to 0$ (meaning that the model is in the AP universality class for any $q>0$), for 
$q\to 1$ (in which case we have OP for any $q<1$), or for some $0<q<1$?

The numerical answer is clear and surprising in its radicalness: For any $q<1$ we find OP,
if we go to large enough $L$, even if $q=0.999$ (see Fig.~\ref{nwrap_L-2Dsq}). More precisely, 
we see that the difference between even and odd $L$ disappears rapidly when $L$ increases, and both 
curves seem to converge to a finite $n_{c,\rm wrap}$ for $L\to\infty$. It seems that even 
the slightest mistake in the agglomeration process completely destroys the phenomenon and places 
the model in the OP universality class.

\begin{figure}
\includegraphics[width=0.5\textwidth]{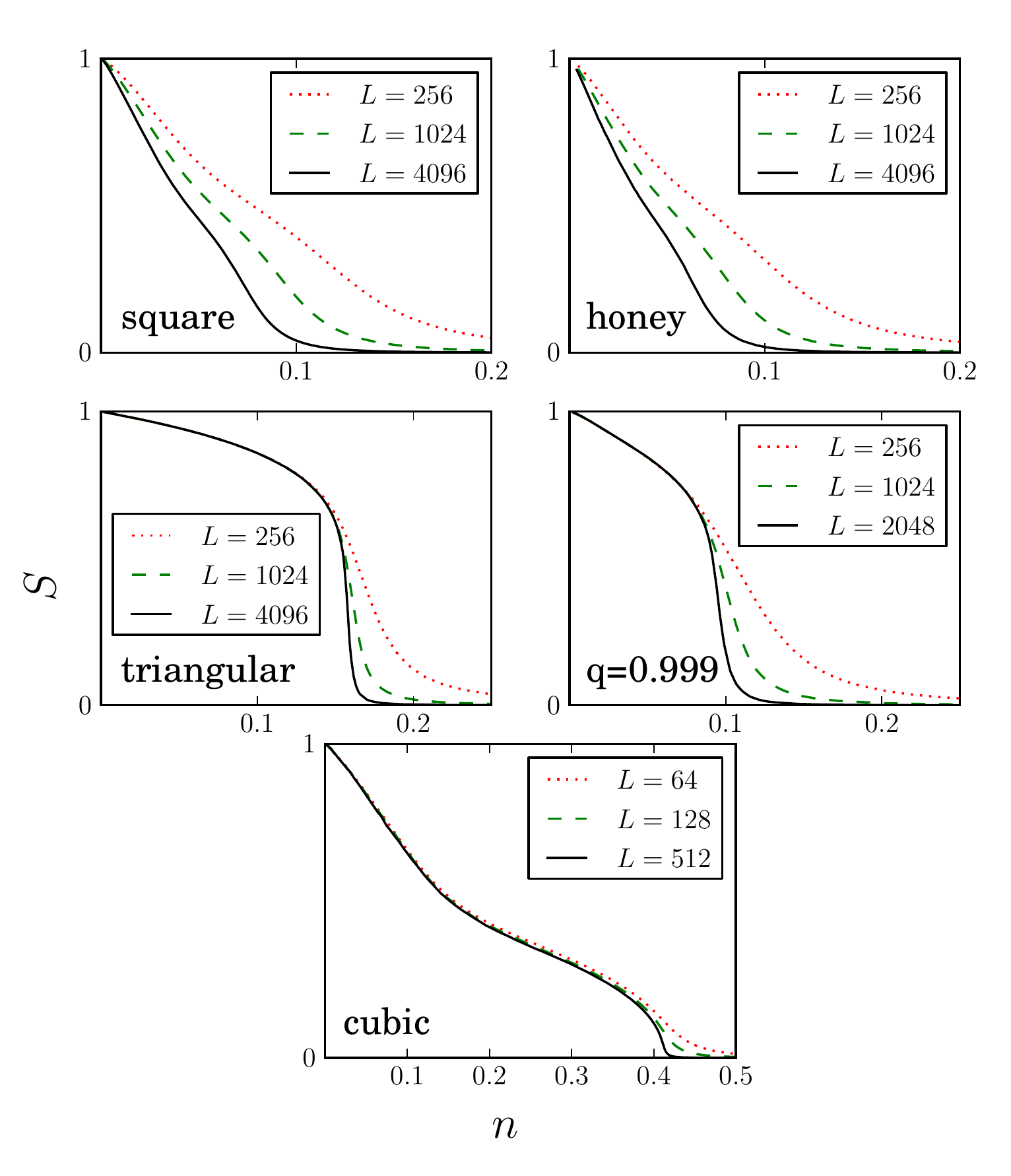}
\caption{\label{smax-lattice} (Color online) The five panels of this figure correspond to 
(i) square lattices with $q=1$, (ii) honeycomb lattices, (iii) triangular lattices, (iv) 
square lattices with $q=0.999$, and (v) cubic lattices. In all cases, periodic b.c. were 
used. In all cases except case (iv), $q=1$.}
\end{figure}

This is confirmed by looking at the order parameter 
\be
    S = \langle m_{\rm max}\rangle / N
\ee
where $m_{\rm max}$ is the size of the largest cluster. For infinite systems, $S=0$ for $n>n_c$
and  $S>0$ for $n<n_c$. For OP, one has $S \sim (n_c-n)^\beta$ for $n$ slightly below $n_c$, and 
the usual finite size scaling (FSS) behavior
\be
    S \sim L^{\beta/\nu} f[(n_c-n) L^{1/\nu}]      \label{FSS}
\ee
for finite systems. In Fig.~\ref{smax-lattice} we show $s$ versus $n$ for various cases, all
with periodic b.c. In addition to two panels for other lattices discussed in later subsections
(``honey", ``cubic") we show results for the triangular lattice and for square lattices with
$q=1$ (``square") and with $q=0.999$. We see that the results for $q<1$ are very similar to 
those for the triangular lattice, while they are completely different from those for $q=1$. 
A data collapse for verifying the FSS ansatz would of course not be perfect, but it seems
that the model with $q<1$ is in the OP universality class, for any $q<1$. 

\subsection{Other regular lattices}

In addition to the square and triangular lattices we now study also the honeycomb lattice as 
a third lattice with $d=2$, and the simple cubic lattice as an example of a 3-d lattice.

\subsubsection{Honeycomb lattice}

Although we measured also other observables (such as the wrapping probabilities), we show here 
only the behavior of the order parameter $S$. As seen in Fig.~\ref{smax-lattice},
the behavior here is very similar to that for the square lattice. In particular, we see no
indication for the FSS ansatz with finite (non-zero) $n$.

\subsubsection{3-d simple cubic lattice} \label{3d}

The behavior on the simple cubic lattice is more subtle. On the one hand, we clearly see in
Fig.~\ref{smax-lattice} an
indication for a non-zero value of $n_c$, with $n_c \approx 0.41$. On the other hand, as 
for the square and honeycomb lattices we see that the slope $dS/dn$ is not monotonic. In all 
three cases the growth of the largest cluster slows down when $S \approx 1/4$, and accelerates
again when $S>1/2$. Alternatively, it seems as if two behaviors are superimposed: For large 
$n$ and $S < 1/3$ it seems as if the curves would extrapolate to $S=1/2$ for $n\to 0$, but then 
(as $n$ decreases further) $S$ rises again sharply, to reach $S=1$. Although this scenario is 
too simplistic, we will see in the next section that it catches some of the relevant physics.

\begin{figure}
\includegraphics[width=0.5\textwidth]{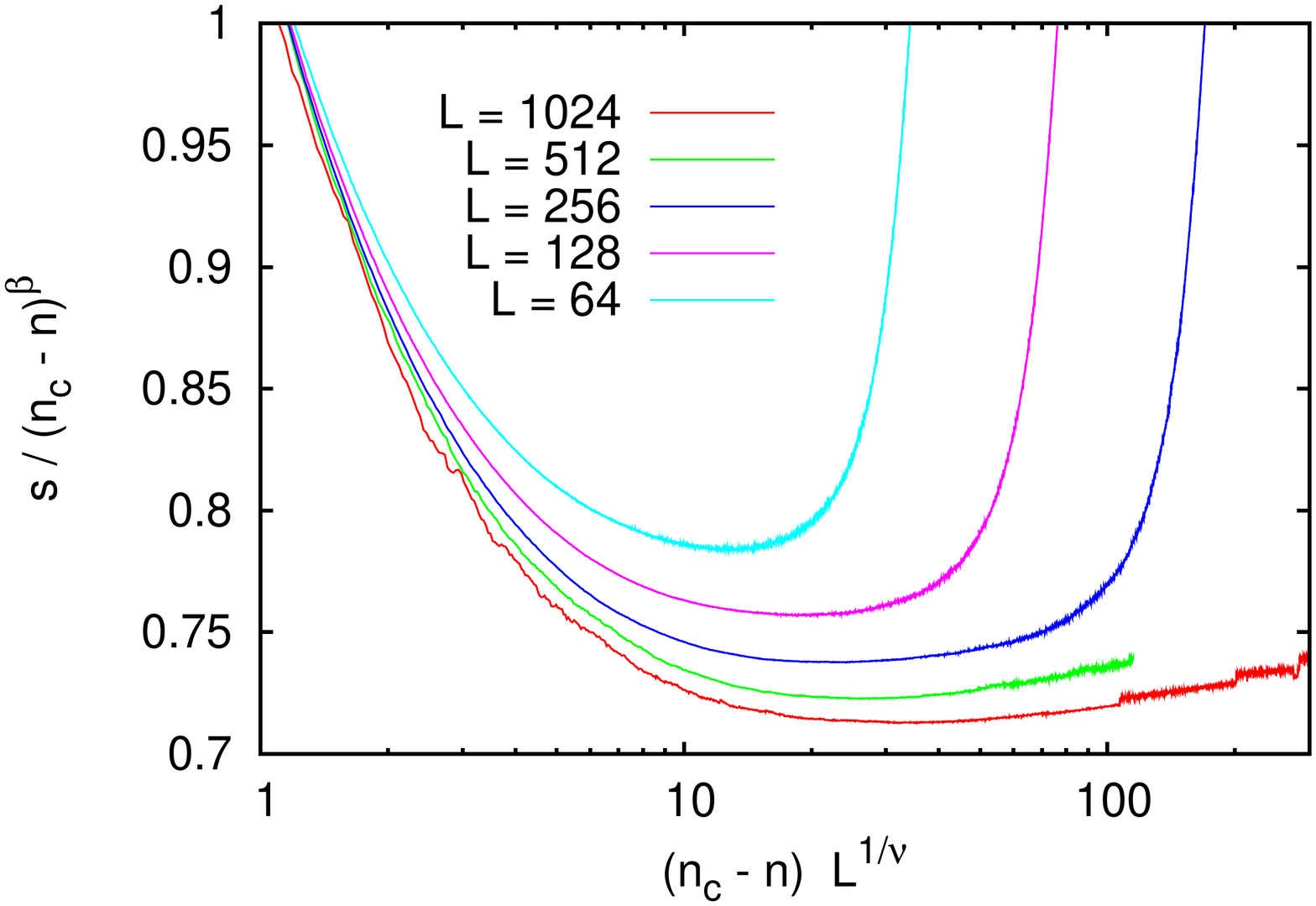}
\caption{\label{smax-3d-beta_0} (Color online) Plot of $S/(n_c-n)^\beta$ against $(n_c-n)L^{1/\nu}$,
using the exponents of ordinary 3-d percolation. According to the FSS ansatz, these curves should 
collapse and should be horizontal in the limit where we first take $L\to\infty$ and then $n\to n_c$.
This seems not to be the case.}
\end{figure}

\begin{figure}
\includegraphics[width=0.5\textwidth]{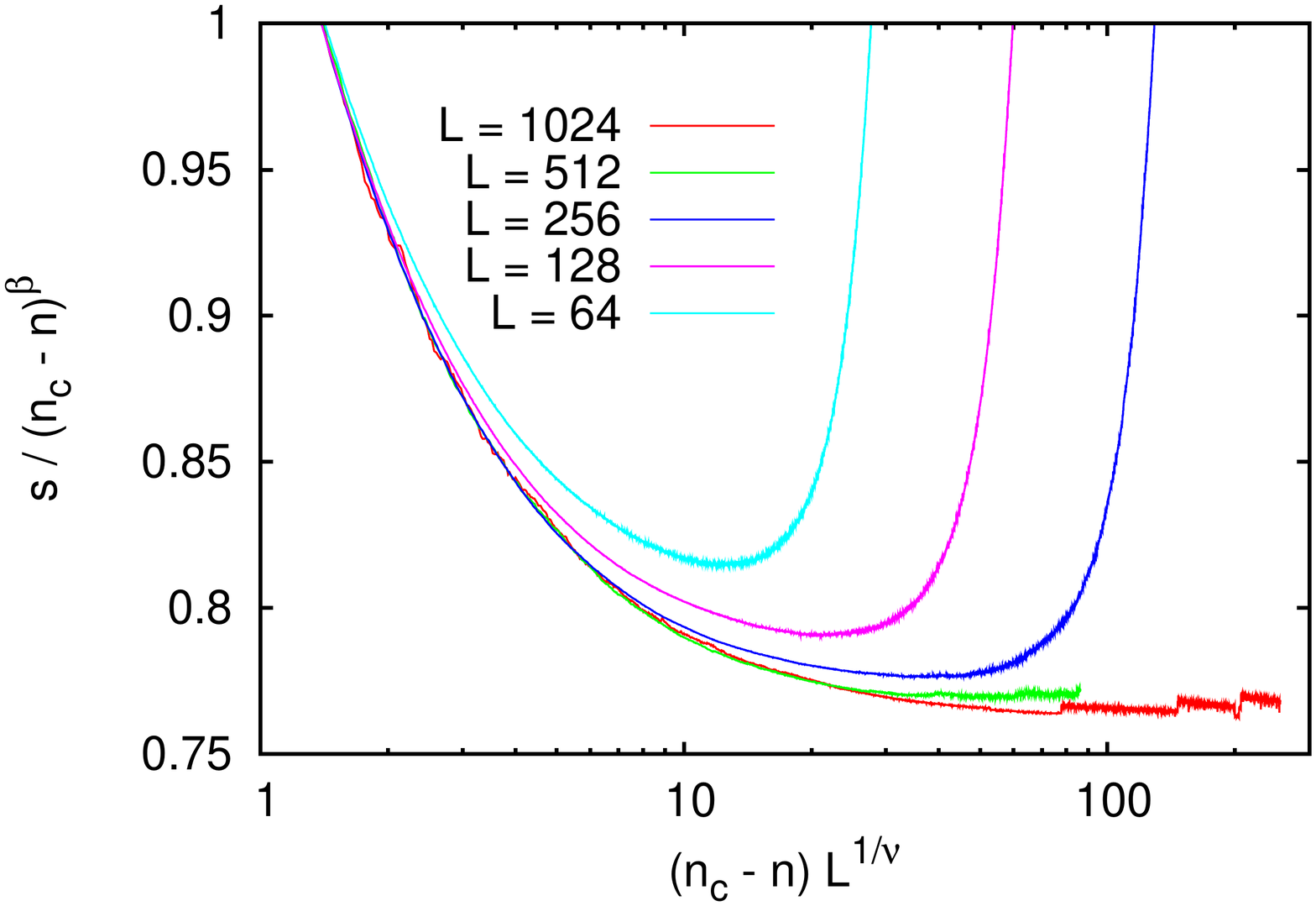}
\caption{\label{smax-3d-beta} (Color online) Analogous to the previous plot, but using $\beta=0.435,
D_f = 2.5220$, and $\nu = 0.91$.}
\end{figure}

\begin{figure}
\includegraphics[width=0.5\textwidth]{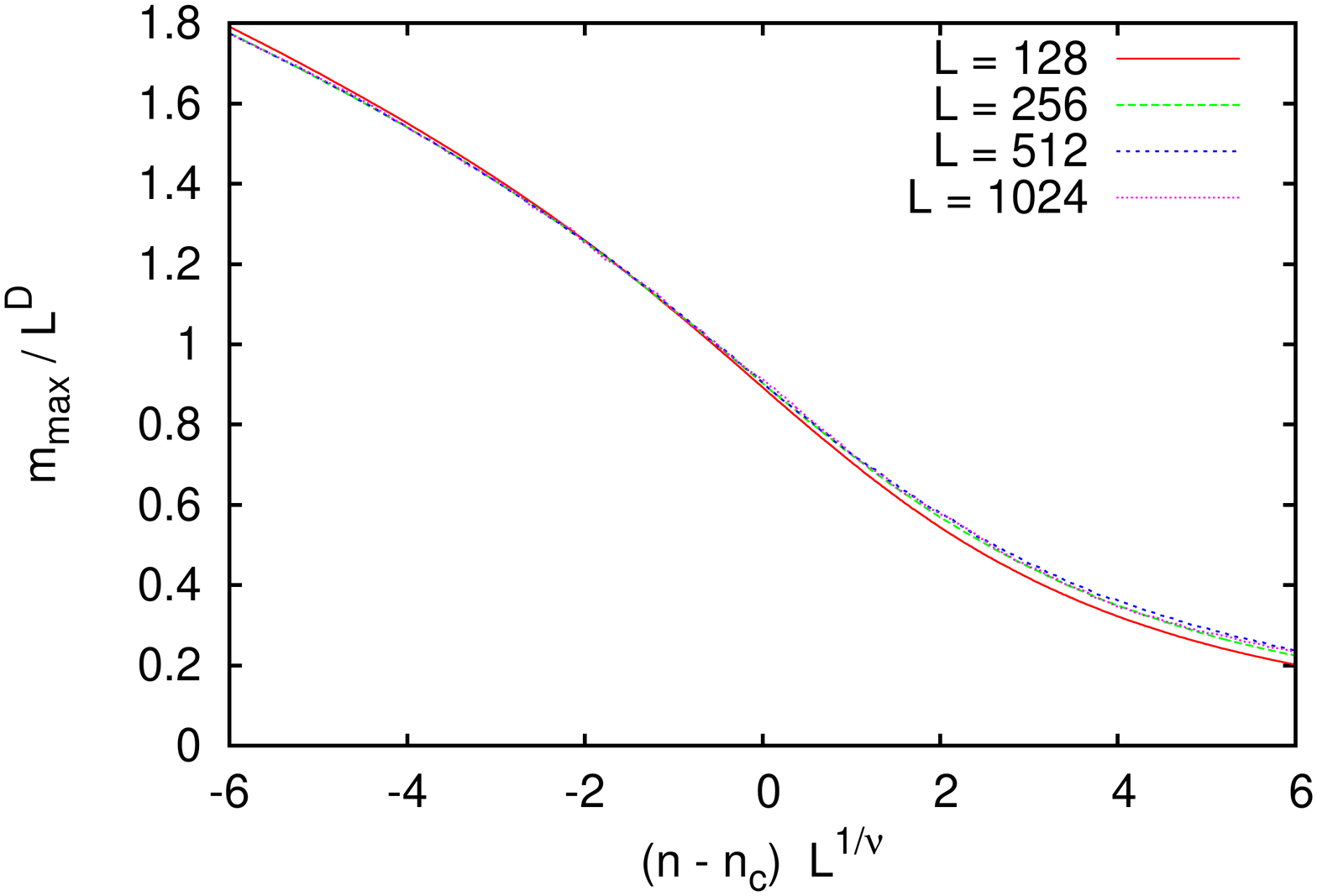}
\caption{\label{smax-3d-collapse} (Color online) Plot of $S L^{\beta/\nu} = m_{\rm max}/L^{D_f}$ versus
$(n-n_c)L^{1/\nu}$ for AP on the simple cubic lattice, using $n_c = 0.411$ and the exponents given in
Eq.~(\ref{3d-expon}).}
\end{figure}

Using the critical exponents for 3-dimensional OP, $\beta = 0.4170(3), D_f = 2.5226(1),$ and
$ \nu = 0.8734(5)$ \cite{Deng-2005}, one obtains an acceptable data collapse when plotting 
$m_{\rm max}/L^{D_f}$ against $(n-n_c)L^{1/\nu}$, with $n_c = 0.411$. But the behavior 
for large $L$ is not given by $S\sim (n-n_c)^\beta$ with the value of $\beta$ given above,
see Fig.~\ref{smax-3d-beta_0}. The latter plot is much improved, if we use instead 
\be
   \beta=0.435,\;\; D_f = 2.522,\;\; \nu = 0.91,  \label{3d-expon}
\ee
together with $n_c = 0.4110$ (see Fig.~\ref{smax-3d-beta}). With these exponents we also obtain 
a good collapse of $m_{\rm max}/L^{D_f}$ against $(n-n_c)L^{1/\nu}$, see Fig.~\ref{smax-3d-collapse}. 
The main deviation from a perfect collapse in this plot is due to the smallest lattice, and is obviously
a finite-size correction to the FSS ansatz. We do not quote error bars for the values in 
Eq.(\ref{3d-expon}), as they are not yet our final estimates.

\section{Bipartivity and Spontaneous symmetry breaking}
    \label{BSSB}

\subsection{Uniqueness of cluster colors}

Our first observation is that infinite square, honeycomb and cubic lattices are bipartite, while
the triangular lattice is not. The next observation is that finite square lattices of size $L\times L$
are still bipartite, if $L$ is even and periodic boundary conditions are used, but global bipartivity 
is lost when either $L$ is odd or helical b.c. are used. These observations strongly suggest that 
it is indeed bipartivity that is responsible for peculiarities of AP on these lattices.

\begin{figure}
\begin{centering}
\includegraphics[width=0.5\textwidth]{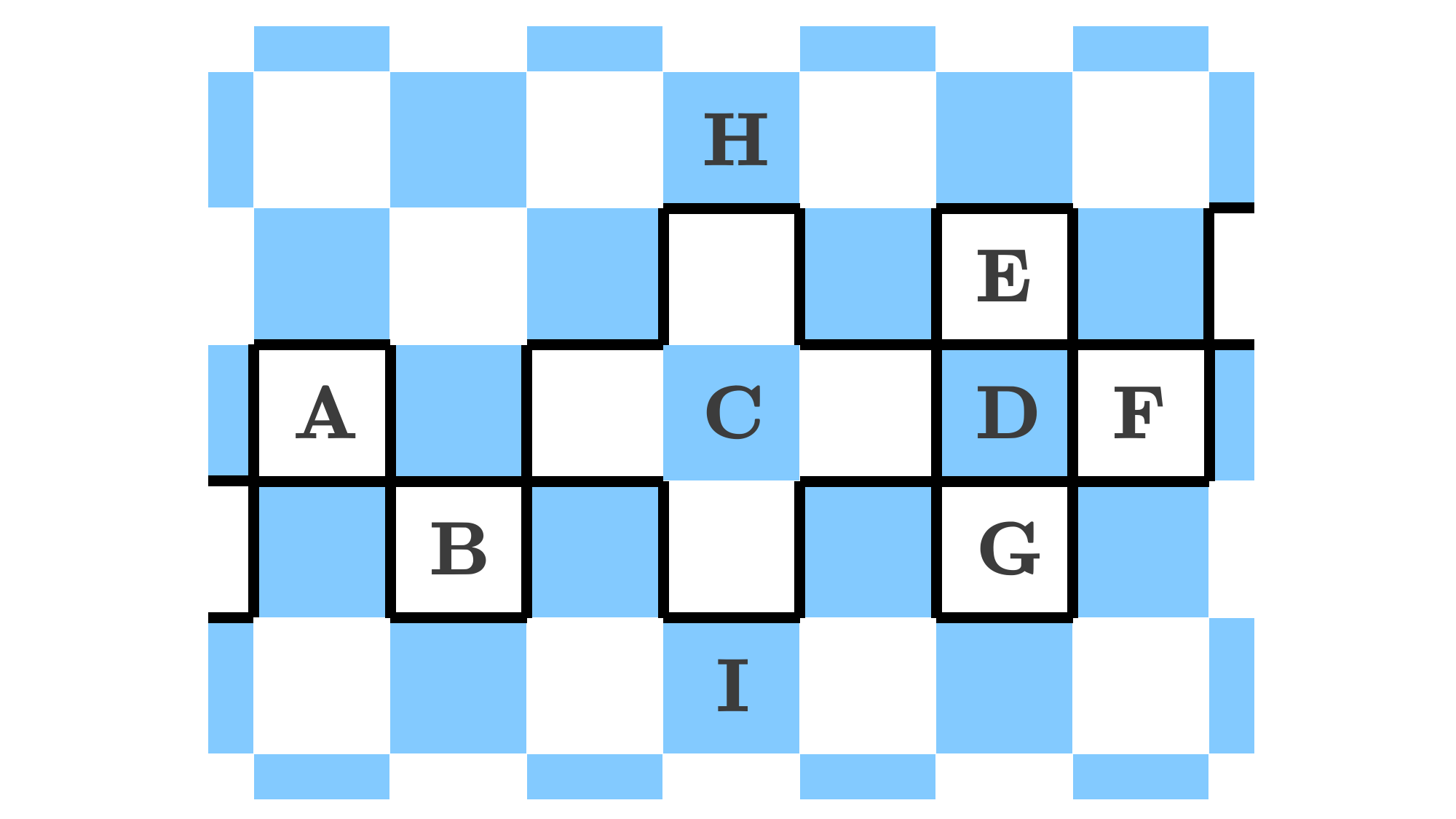}
\par\end{centering}

\caption{\label{illustration-bipartite} (Color online) Part of a square lattice with cluster boundaries 
indicated by black lines. Plaquettes correspond to nodes of the graph. Nine clusters are labeled with 
letters $A-I$. Six of them ($A,B,D-F$) are single nodes, one ($C$) has five nodes, and two ($H,I$) are very large.
Three of them ($D,H,I$) are blue, the other six are white. If cluster $D$ is chosen as target, it
merges with $C,E,F$, and $G$ and the new cluster is white. If, however, $F$ is chosen as target, the new blue
cluster would consist of $D, H$, and $I$.}
\end{figure}

In a bipartite graph, to each node can be assigned one of two colors. We now show that this is 
extended from single {\it nodes} to arbitrarily large {\it clusters}, if the rules of AP are strictly
followed. Before we do this, we need two definitions:

{\it Definition:} {\sl The {\it surface} of a cluster $C$ is the set of all nodes in $C$ that have at 
least one link to a node not contained in $C$.}

{\it Definition:} {\sl If all surface nodes in $C$ have the same color, then we say that $C$ also has this 
color. Otherwise, the color of $C$ is not defined.}
 
We can now prove the following {\it Theorem:}\\
{\sl (i) If clusters are grown according to the AP rules on a bipartite network, they always have a 
well defined color.\\
    (ii) All neighbors of a given cluster have the opposite color.\\
    (iii) If a target of color $c$ is chosen for agglomerating all its neighbors, the new cluster
has the opposite color ${\bar c}$.}

For an illustration see Fig.~\ref{illustration-bipartite}.

Proof: The proof follows by induction. First, the theorem is obviously true for the starting configuration,
where all clusters are single nodes. Then, let us assume it is true for all agglomeration steps up to 
(and including) step $t$. Let us call $c$ the color of the target cluster at step $t+1$, and ${\bar c}$ the
opposite color. Then all neighbors of the target have color ${\bar c}$, so that after joining them
the new cluster also has color ${\bar c}$, proving thereby (i) and (ii). On the other hand, all neighbors 
of the neighbors had color $c$, and these form the neighbors of the new cluster, which proves (iii).
\hfill
$\blacksquare$


Notice that it was crucial for the proof that {\it all} neighbors of the target were joined, so that none 
of the neighbors of the target is a neighbor of the new cluster. This shows why imperfect agglomeration as 
considered in Sec.~\ref{finite_q} leads to situations where the theorem does not hold.

\subsection{Coexistence of large clusters}

\begin{figure}
\begin{centering}
\includegraphics[width=0.5\textwidth]{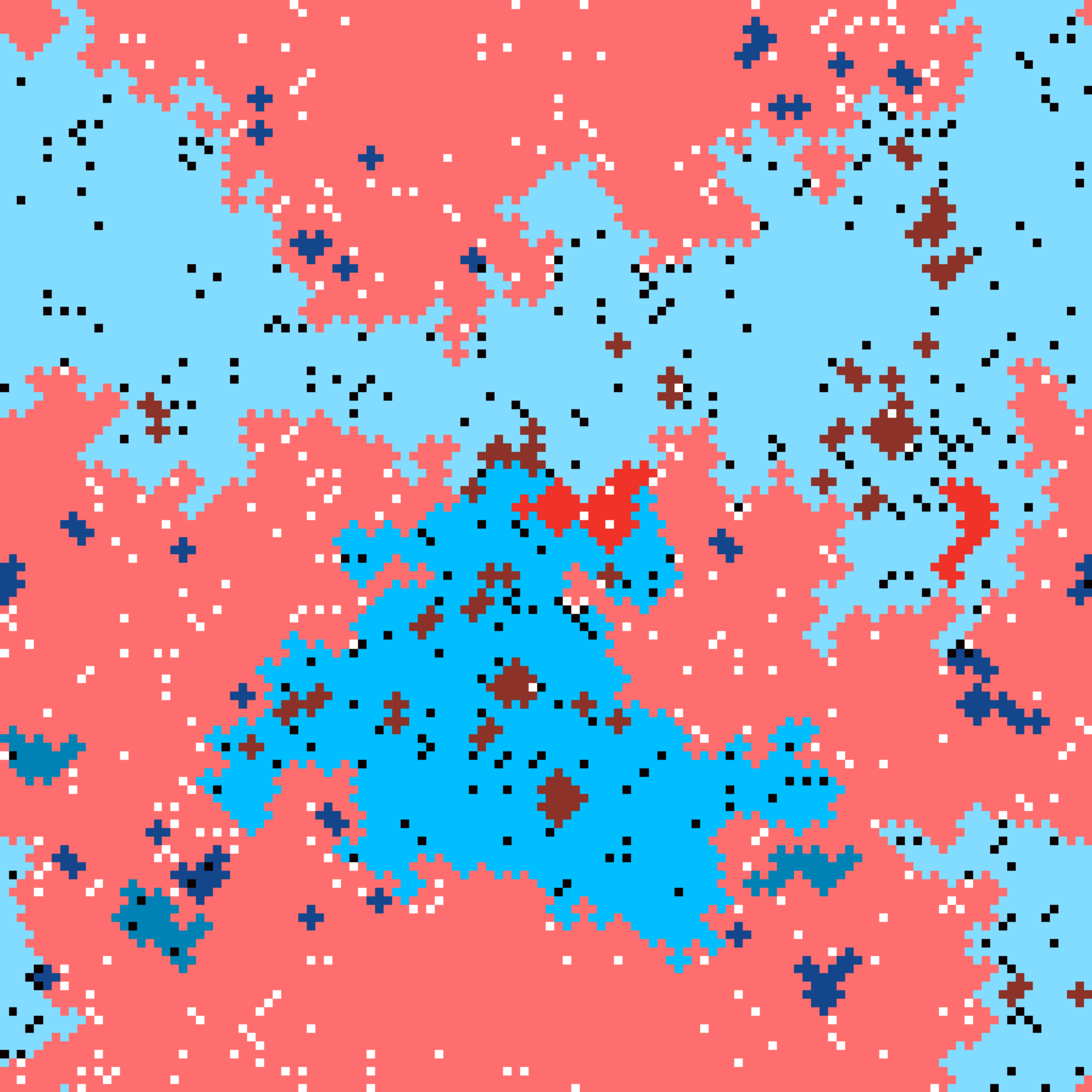}
\par\end{centering}
\caption{\label{square-config} (Color online) A configuration
with three large clusters at $n\approx0.03$ and $L=128$. For clusters of mass $>1$ the two colors are 
red and blue, with the larger clusters more bright and the smaller ones more dark. ``Blue" singletons
are colored white for better visibility and in order to distinguish them from ``red" singletons which 
actually are indicated by black squares. Notice that no two clusters of the same color ever touch in
this figure. Wherever they {\it seem} to touch, there is indeed a small cluster of the opposite color 
intervening. In spite of the size of the largest clusters, none of them has yet wrapped in the vertical
direction.}
\end{figure}

A typical configuration on a $128\times 128$ lattice with three large clusters, none of which has yet 
wrapped vertically, is shown in Fig.~\ref{square-config}. Such a configuration would have an astronomically 
small probability in OP, since in OP the chance is very small to have more than one large cluster. If there
were two large clusters at any time, they would immediately merge with very high probability.
Obviously, in AP there exists a mechanism that prevents clusters of opposite color to merge fast, leading 
to the coexistence of large clusters of opposite colors.

This looks at first paradoxical. Take the two largest clusters in Fig.~\ref{square-config}. If either of 
them were chosen as target, they would merge immediately. Why should this not happen? The crucial point is 
that each cluster is chosen as target with the same probability, and there are many more small clusters 
than large ones. The chances are thus overwhelming that neither of the large clusters is chosen as target, but 
a small cluster is picked instead.
But in that case the two largest clusters cannot merge, because they have opposite color and all 
neighbors to be merged must have the same color. Thus its is most likely that a random agglomeration 
step merges one small cluster of color $c$ with several (small and large) clusters of color $\bar c$.

\begin{figure}
\begin{centering}
\includegraphics[width=0.5\textwidth]{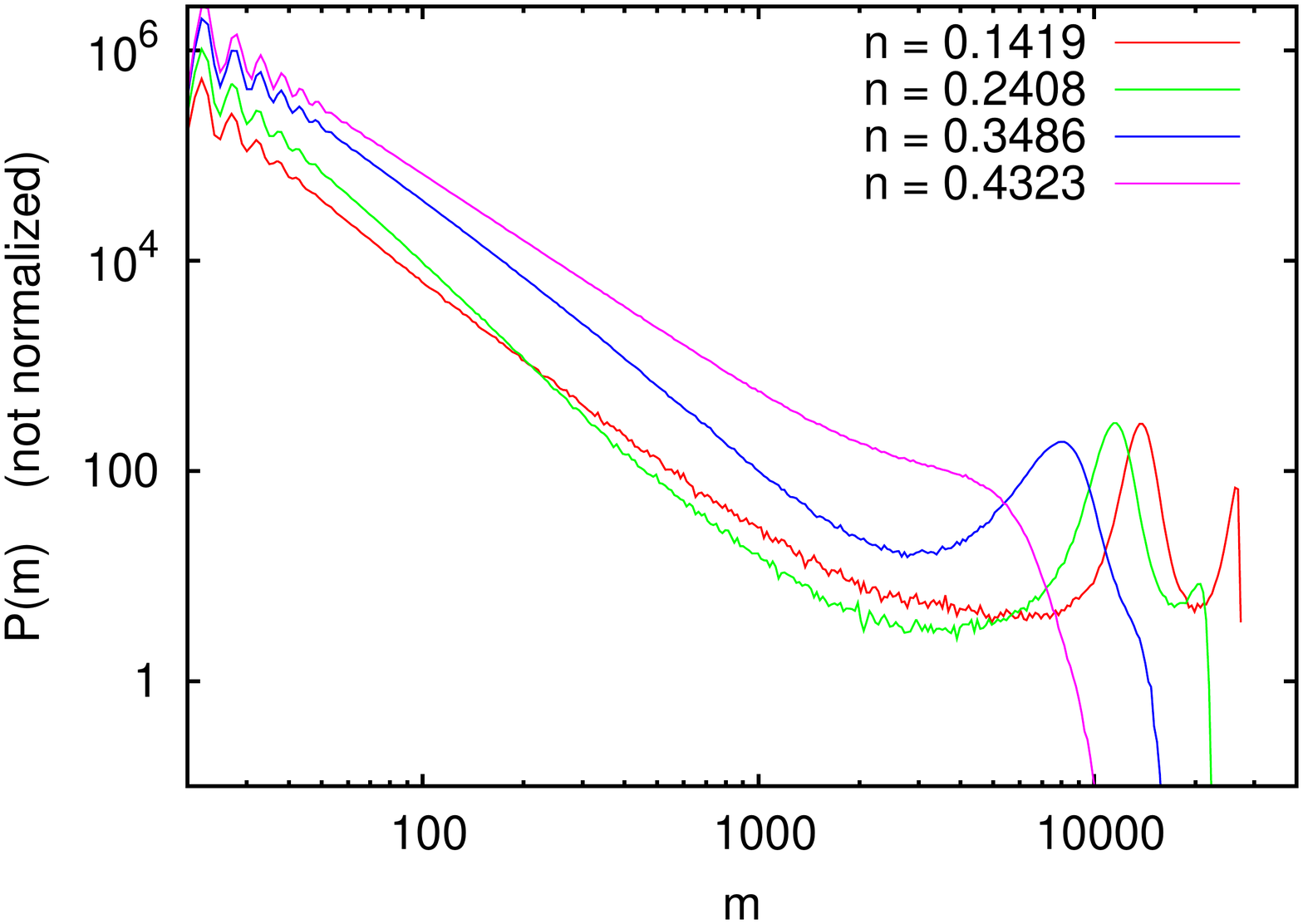}
\par\end{centering}
\caption{\label{3d-massdistr} (Color online) Cluster mass distributions for simple cubic lattices 
with $L=32$. The curve for $n=0.4323$ is subcritical, while the other curves are supercritical. In 
contrast to the case of OP, where the mass distribution develops a single peak in the supercritical 
phase, now (i.e. for AP) we see two peaks. They correspond to clusters of opposite colors.}
\end{figure}

\begin{figure}
\begin{centering}
\includegraphics[width=0.5\textwidth]{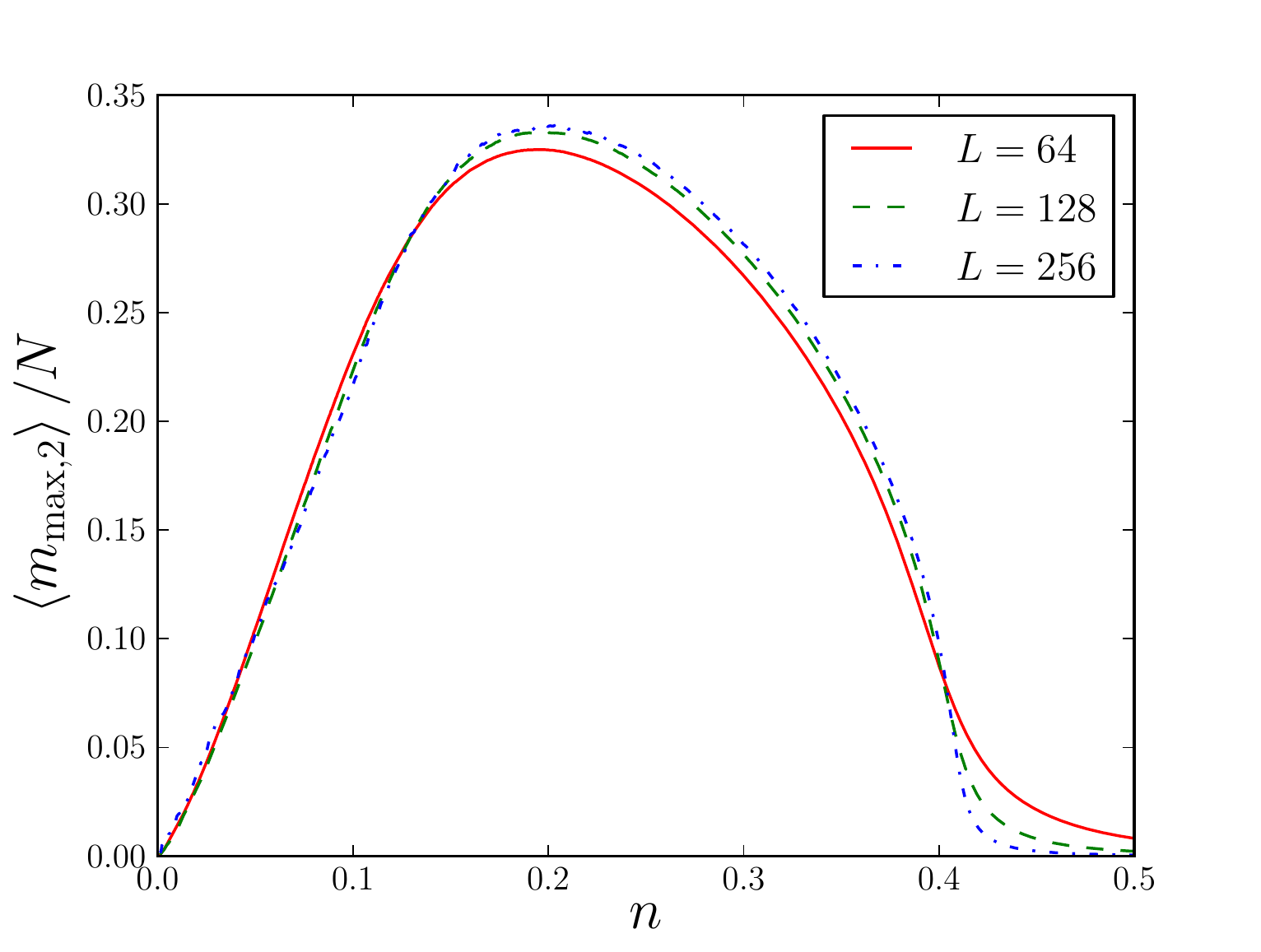}
\par\end{centering}

\caption{\label{smax2_cubic} Average size of the second largest cluster on simple cubic lattices 
with even size. In contrast to OP, where $m_{\rm max,2}$ peaks near the percolation transition 
and decreases fast when one goes into the supercritical phase, here the second largest cluster 
continues to grow far beyond the percolation transition $n_{c}\approx0.411$.}
\end{figure}

In two dimensions this means also that the two large clusters of opposite color prevent each other from 
wrapping. In three dimensions this is not the case. Thus AP is in three dimensions more similar to OP,
although it still should show several large clusters in the critical and supercritical regimes. To test
this prediction we show two figures. 
Fig.~\ref{3d-massdistr} shows that mass distributions in the supercritical phase have two peaks (in
contrast to OP), corresponding to the fact that AP on bipartite graphs has {\it two} giant clusters 
of opposite colors. The same conclusion is drawn from Fig.~\ref{smax2_cubic}, where we show the 
average normalized size $m_{\rm max,2}$ of the second largest cluster as a function of $n$. We see that 
$m_{\rm max,2}$ starts to increase at $n_c$ and continues to grow as one goes deeper into the 
supercritical phase, while it would peak at $n_c$ in OP.
\subsection{Surface color statistics}

\begin{figure}
\begin{centering}
\includegraphics[width=0.5\textwidth]{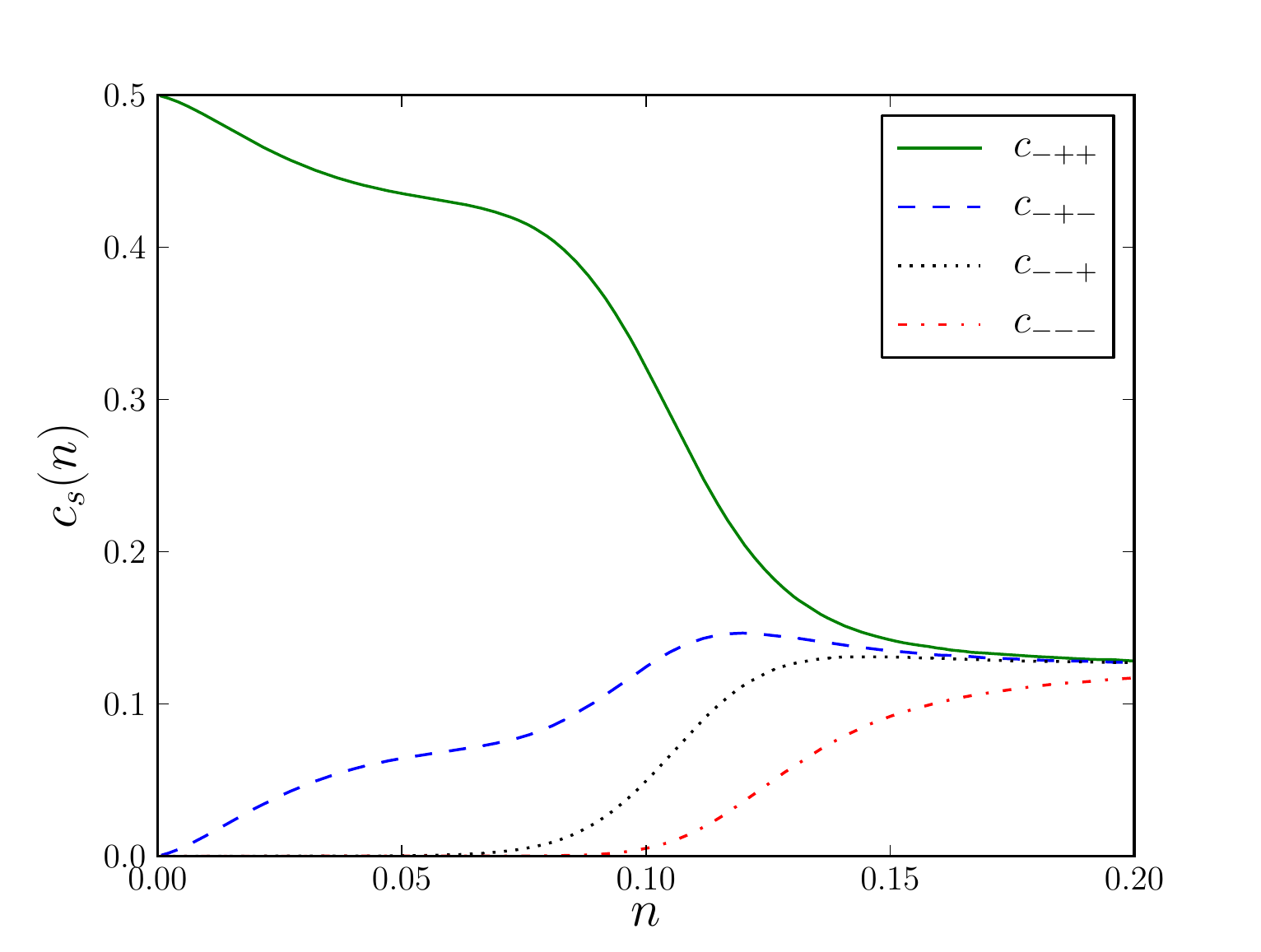}
\par\end{centering}
\caption{\label{signstat3_2D} (Color online) Probabilities $c_{s}(n)$ of color patterns 
$s = (-++),\;(-+-),\;(--+),$ and $(---)$ for the largest three clusters, plotted against $n$. 
The first index (here always `-') gives the color of the largest cluster, while the other two 
are for the second and third largest. The data are for square lattices of size $L=512$.}
\end{figure}

While these two figures show that there is indeed more than one giant cluster in AP on bipartite 
lattices, they do not yet prove that these clusters have opposite colors. To verify also this 
prediction we denote the two colors as `+' and `$-$', and define $c_{ijk\ldots}$ ($i,j,k\ldots \in\{+,-\}$) 
as the probabilities that the largest cluster has color $i$, the second largest $j$, etc.
These probabilities are normalized such that $\sum_{ijk\ldots} c_{ijk\ldots} = 1$. Due to the 
symmetry under exchange of colors, $c_{ijk\ldots}=c_{{\bar i}{\bar j}{\bar k}\ldots}$.
In Fig.~\ref{signstat3_2D} we plot the four probabilities $c_{-jk}$ for square lattices with $L=512$
against $n$. While they are all equal to $\approx 1/8$ for large $n$, this degeneracy is lifted as the 
agglomeration process proceeds. The most likely color pattern is $(-++)$, followed by $(-+-)$. Both 
have opposite colors for the two largest clusters. The 
least likely pattern has all colors the same.

\begin{figure}
\begin{centering}
\includegraphics[width=0.5\textwidth]{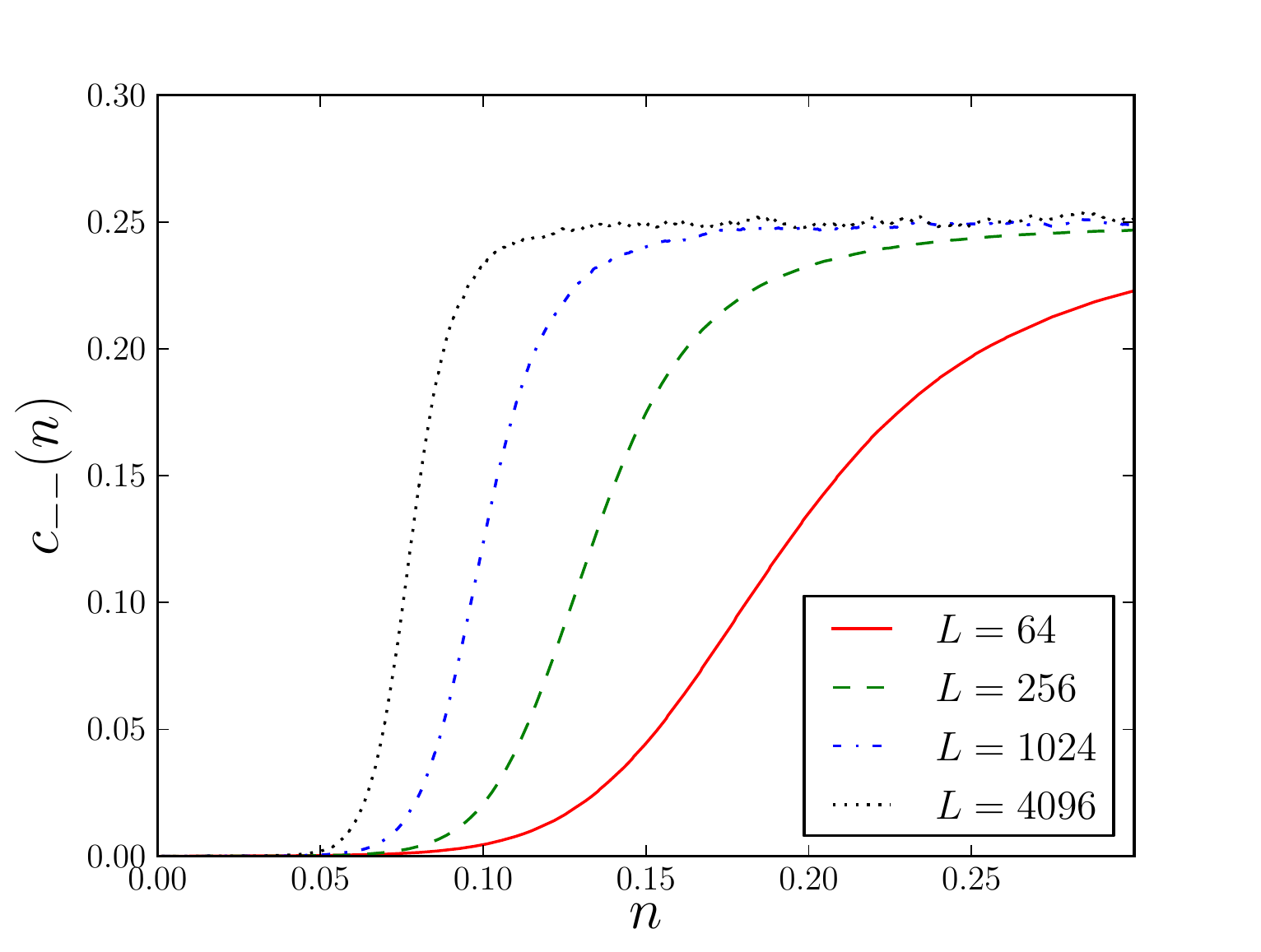}
\par\end{centering}
\begin{centering}
\includegraphics[width=0.5\textwidth]{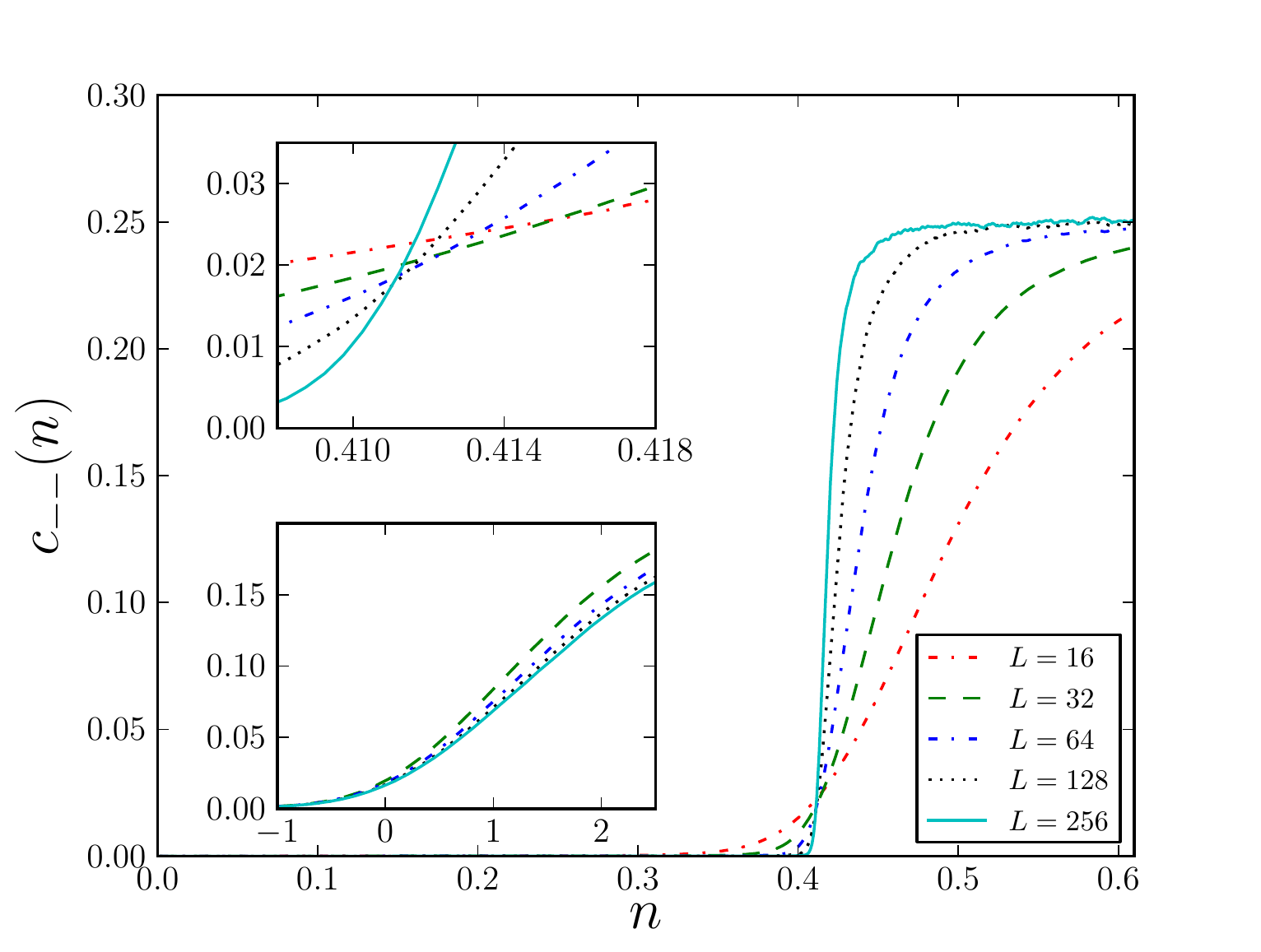}
\par\end{centering}
\caption{\label{signstat2_L_all} (Color online) Probabilities that the two largest clusters have the same color.
According to our theory, these probabilities should vanish in the supercritical phase, if $L\to\infty$.
Panel (a) is for the square lattice, panel (b) for the cubic. The upper inset in panel (b) shows the region 
close to the critical point. The lower inset shows a data collapse plot, $c_{--}(n)$ against
$(n-n_c) L^{1/\nu}$ with $n_c = 0.4109$ and $\nu = 1.01$.}
\end{figure}

In Fig.~\ref{signstat2_L_all} we show how the $n$-dependence of the probability $c_{--}$ that the 
two largest clusters have the same color changes with system size $L$, both for 2 and 3 dimensions.
There is a dramatic difference: While the data support our conclusion that there is no transition 
at any finite $n$ in case of the square lattice (the effective transition point moves to zero as 
$L$ increases), there is a clear indication for $n_c=0.411$ in case of the cubic lattice.

More precisely, the lower inset in Fig.~\ref{signstat2_L_all} shows a nearly perfect data collapse
when plotting $c_{--}(n)$ against $(n-n_c) L^{1/\nu}$, with $n_c = 0.4109$ and $\nu = 1.01$. The latter 
values are very close to the values obtained in Sec.~\ref{3d} from the data collapse for the ordinary 
order parameter, but sufficiently far from them to call for further, so far unnoticed, corrections to 
scaling. Combining both sets of parameters, accounting for such corrections by increasing the error
estimates, and noticing that the system sizes in Fig.~\ref{signstat2_L_all} are much smaller than those 
in Figs.~\ref{smax-3d-beta_0} to \ref{smax-3d-collapse}, we get our final result
\be
   \beta = 0.437(6),\;\; D_f = 2.523(3),\;\; \nu = 0.918(13), 
\ee
and $n_c = 0.4110(1)$. 

Since the differences between these exponents and those of OP are about three to four error bars, we 
conjecture that the two models are not in the same universality class. But more studies are 
needed to settle this question beyond reasonable doubts. In any case, Fig.~\ref{signstat2_L_all}
should leave no doubt that $1/4 - c_{--}(n)$ is as good an order parameter for the symmetry breaking 
aspects of the transition, as $S$ is for the percolation aspects.

\subsection{Lattices with local bipartite structure}

Let us finally discuss the case where we have locally a bipartite lattice, but where global bipartivity
is broken by the boundary conditions. In that case the boundary conditions are irrelevant as long as the 
cluster does not wrap around the lattice. In particular, we expect that such a system is 
not in the OP universality class, if the globally bipartite system is not either. More precisely, we 
expect that clusters of size $<L$ are unaffected by the boundary conditions. Whether critical exponents
like the order parameter exponent $\beta$ are affected, which are defined through the behavior of the 
supercritical phase, is an open question. 
 
\section{Random bipartite graphs}

One minor problem in simulations of random bipartite networks is that we want connected graphs, but 
the most straightforward way of generating it leads to graphs that are not connected. We thus start 
with $N^*$ nodes, divide them into two equally large groups, and add $zN^*/2$ edges which have the 
two ends in different groups. Here $z$ is the average degree of the entire graph, which is chosen
as $z = 2$ in the following.

\begin{figure}
\begin{centering}
\includegraphics[width=0.5\textwidth]{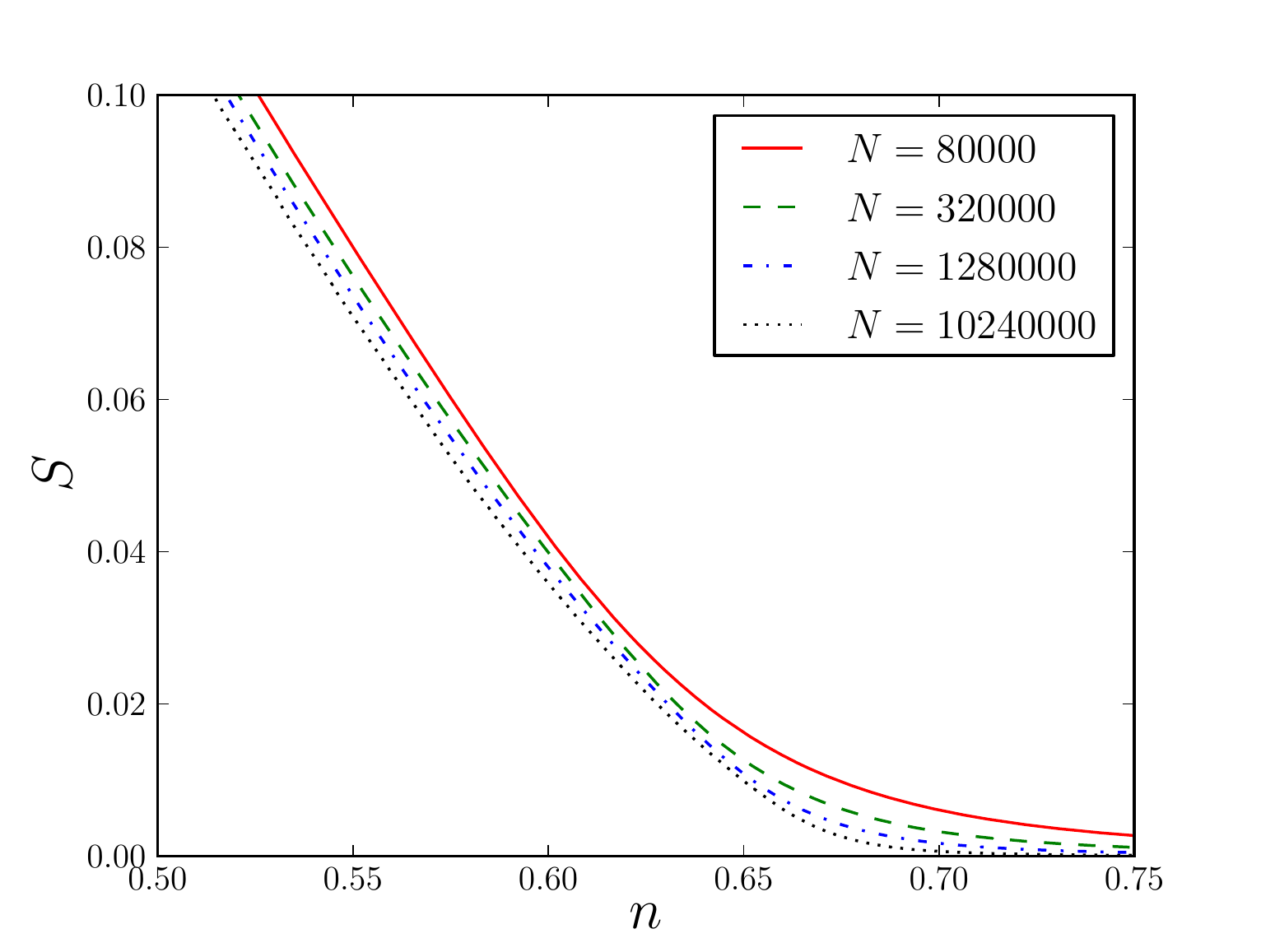}
\par\end{centering}
\caption{\label{fig:AP_smax-bipart} Fraction of nodes in the largest cluster
$\left\langle m_{\mathrm{max}}\right\rangle /N$ for random bipartite
networks.}
\end{figure}

\begin{figure}
\begin{centering}
\includegraphics[width=0.5\textwidth]{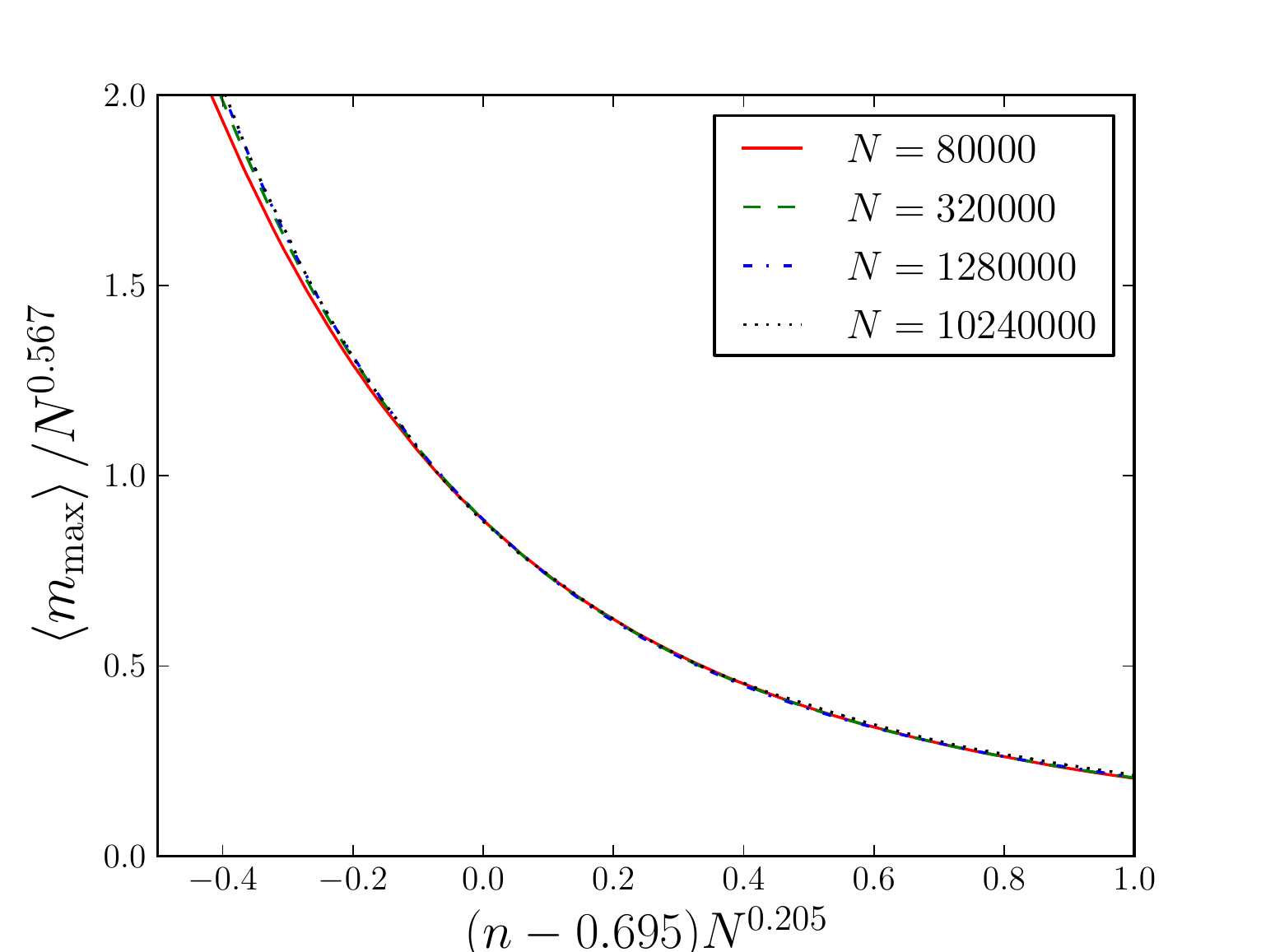}
\par\end{centering}
\caption{\label{fig:AP_smax-bipart-collapse} Scaling collapse for $\left\langle m_{\mathrm{max}}\right\rangle $
in the critical region for random bipartite networks.}
\end{figure}

For this value of $z$, the largest connected component of the network constructed this way has $\approx 
0.7968 N^*$ nodes. If we want to have a connected graph with $N$ nodes, we take $N^*=N/0.7968$ and 
discard all those graphs for which the size of the largest connected component is outside the range 
$N\pm0.01\%$, and for which any of the two components has size outside the range $N/2\pm0.01\%$.

The $n-$dependence of the size of the largest cluster is shown in Fig.~\ref{fig:AP_smax-bipart}.
We assume again a FSS ansatz analogous to Eq.~(\ref{FSS}), with $L$ replaced by $N$ (now $D$ can
of course not be interpreted as dimension, and $\nu$ no longer is a correlation length exponent).
The critical point and the critical exponents are found as (see Fig. \ref{fig:AP_smax-bipart-collapse}):
$n_c=0.695,\;\; \nu=4.88,\;\;D=0.567$. 

\begin{figure}
\begin{centering}
\includegraphics[width=0.5\textwidth]{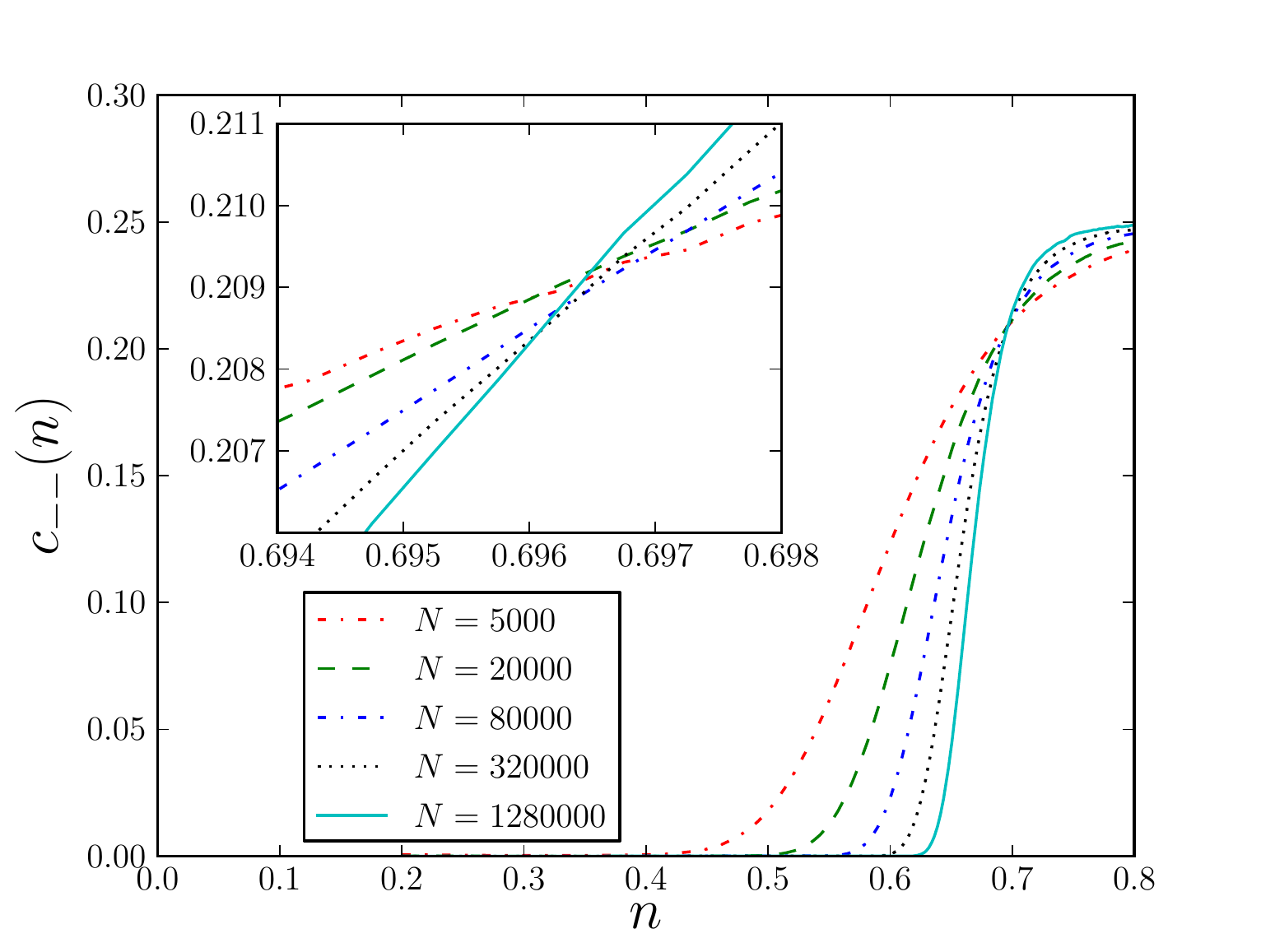}
\par\end{centering}
\caption{\label{fig:signstat2_L_bipart} Probabilities $c_{--}(n)$ for the largest two clusters in a 
random bipartite network to have the same color, for different system sizes. As also seen from the 
inset, these curves cross near the estimated critical point $n_c\approx 0.696$.}
\end{figure}

The probabilities $c_{--}(n)$ for both largest cluster to have the same color are shown in 
Fig.~\ref{fig:signstat2_L_bipart}. As in the 3-d case (Fig.~\ref{signstat2_L_all}b) we see that 
the curves for different system sizes cross exactly at the critical point, suggesting again that 
$c_{--}(n)=0$ in the supercritical phase in the large system limit \footnote{For very small and very 
large $N$, $c_{--}(n)$ is ill defined, because in these limits there can be more than one largest 
(or second largest) cluster, but this affects only regions infinitesimally close to $n=0$ and $n=1$ in the limit $N\to\infty$.}.
We also obtain a perfect data collapse if we plot $c_{--}(n)$ against $(n-n_c)N^{1/\nu}$, with 
slightly different parameters $n_c=0.696,\; \nu = 4.60$ (data not shown). Our best estimates for the 
critical parameters are the compromise 
\begin{equation}
n_{c}=0.695(2),\;\; \nu=4.7(2),\;\;D=0.567(6) \;.
\end{equation}
The values for $\nu$ and $D$ are very close to those for Erd\"os-Renyi networks ($\nu=4.44,\;D=0.60$), 
although they differ by more than one standard deviation. As in the 3-d case, more work would be needed 
to determine whether these differences are significant.

\section{Generalizations to $k-$partite graphs}  \label{k-partite}

A graph is $k-$partite for any $k\geq 2$, if the set of nodes can be divided into $k$ non-empty
disjoint subsets ${\cal N}_m,\; m=1\ldots k$ such that there are no links within any of the ${\cal N}_m$.
As we saw in Sec.~\ref{BSSB}, the appearance of novel structures in AP on bipartite graphs 
depended on the fact that AP does not ``mix" colors: After each agglomeration step, one can 
still associate a unique color to the new cluster. This is no longer true on $k-$partite graphs
with $k>2$. Assume a node $i$ has neighbors with two different colors, $c_1$ and $c_2$ say. Then, 
if $i$ is chosen as a target, the new cluster will display both colors on its surface.

\begin{figure}[t]
\begin{centering}
\includegraphics[width=0.5\textwidth]{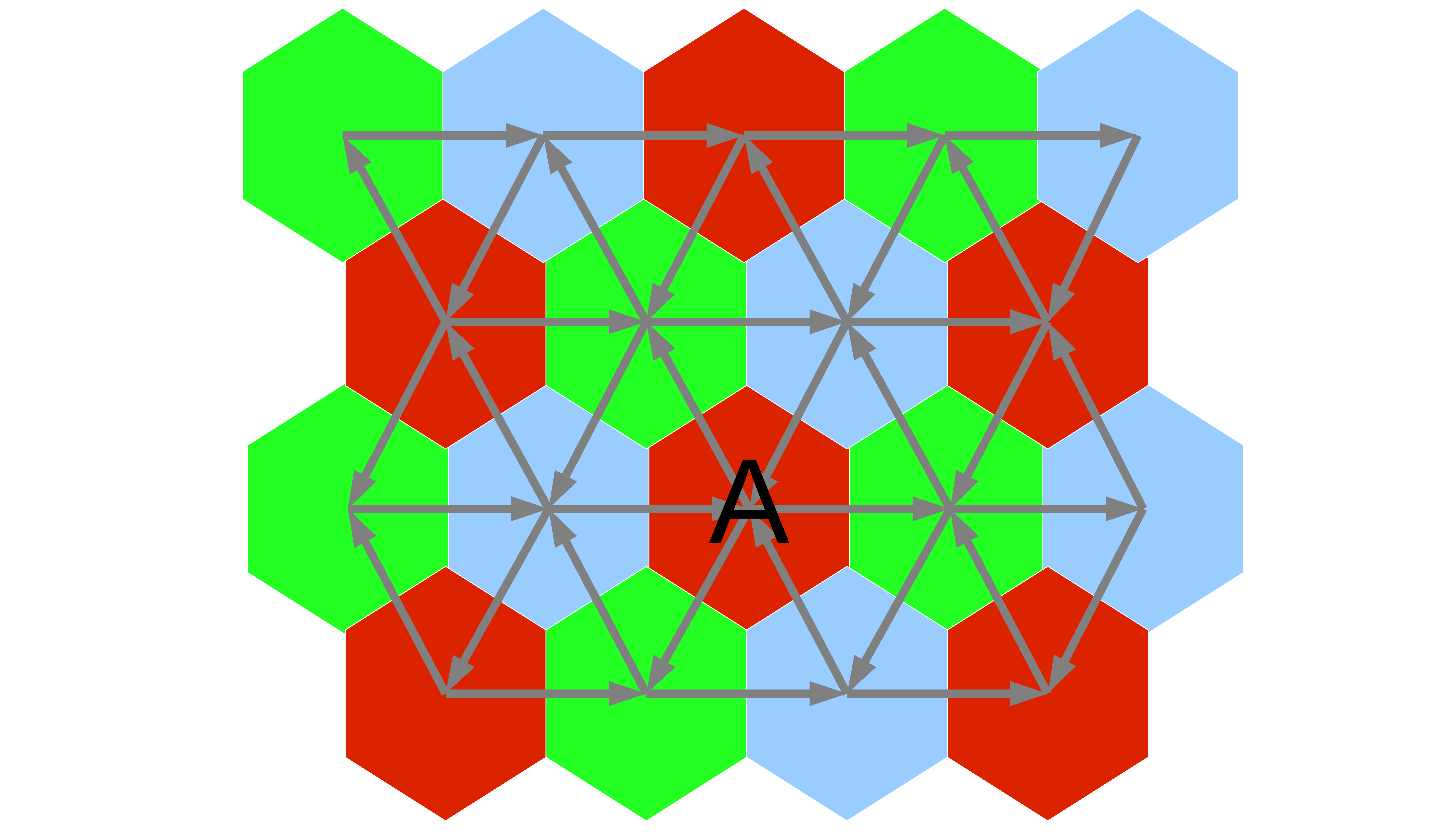}
\par\end{centering}
\caption{(Color online) Part of a a triangular lattice where sites
(hexagons) are colored red, green and blue. The colors are arranged such that no two sites 
with the same color are adjacent, i.e. if neighbors are connected by bonds the lattice is
tripartite. A modified AP process is defined such that a target with color R can join only 
with neighbors of color G, G can join only with B, and B can join only with R. This is 
indicated by the arrows and is denoted by $R\to G\to B\to R$. When node $A$ is chosen as target,
it agglomerates with all G neighbors and becomes itself G, so that the new cluster is all G
on its surface.} \label{RGB-triangle}
\end{figure}

\begin{figure}
\begin{centering}
\includegraphics[width=0.5\textwidth]{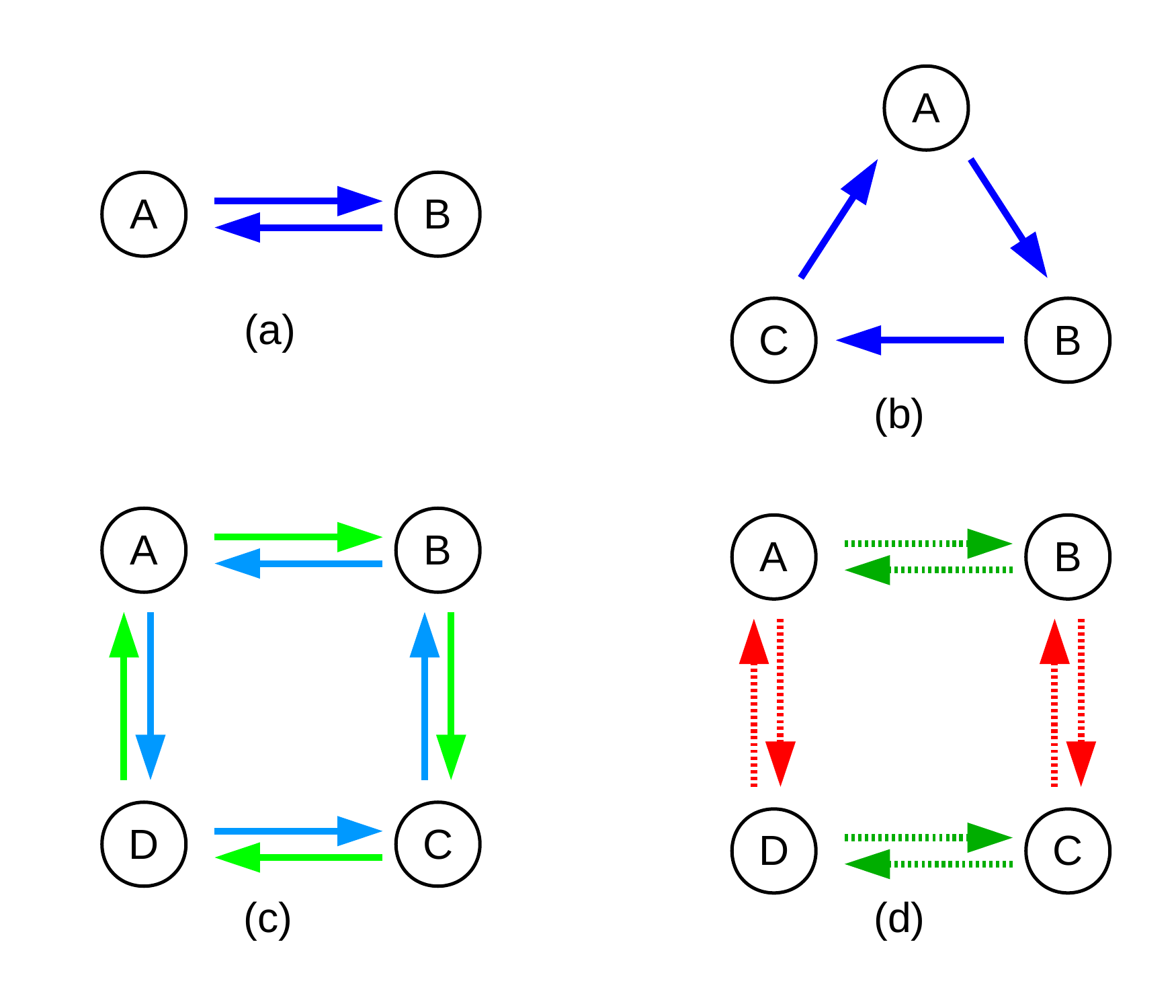}
\par\end{centering}
\caption{\label{AP-rules} (Color online) Each circle represents a color, i.e. an element of 
a partition of a $k-$partite network. An arrow from partition $A$ to partition $B$ means that
a target with color $A$ joins with all nodes of color $B$. Solid arrows indicate cycles that 
are followed at each agglomeration step, while dotted arrows indicate random AP rules where 
different colors are chosen at different time steps.}
\end{figure}

In order to arrive at non-trivial structures we have to generalize the AP rule. Assume we have
a $k-$partite graph with colors $c_1,\ldots c_k$. In Fig.~\ref{RGB-triangle} we show the triangular 
lattice as an example of a tripartite graph with colors red (R), green (G) and blue (B). We 
define then a {\it cycle} $\cal C$ in the set of colors as a closed non-intersecting directed path
$c_{i_1}\to c_{i_2}\to \ldots c_{i_k}\to c_{i_1}$. For tripartite graphs as in Fig.~\ref{RGB-triangle} 
there are two possible cycles, $R\to G\to B\to R$ and $R\to B\to G\to R$ (up to circular shifts).
For each cycle $\cal C$ we define a modified AP rule AP$_{\cal C}$ such that a target with color 
$c$ joins with all neighbors having the color that follows $c$ in $\cal C$, and only those. After
that, the target is recolored to $c$, so that the new cluster has a unique color.

Alternatively, we can define a randomized rule AP$_{\rm random}$ such that each target node $i$
chooses at random a color $c$ (different from its own) and joins with all neighbors of color $c$.
Obviously even more possibilities exist when $k>3$. For instance we can choose the joined neighbors
by following some subset of cycles. Different possibilities are illustrated in Fig.~\ref{AP-rules}.

We have not made any simulations, but we expect a rich variety of different behaviors resulting 
from different rules. It is not clear that in each case AP differs from OP in critical behavior. It 
is also not clear what happens, if one of the partitions of the network is finite. Naively one should 
expect that such finite components should not have any influence on critical behavior (which deals
only with infinite clusters). But the example of finite $q$ in subsection \ref{finite_q} might 
suggest otherwise: It could be that even a small number of nodes that do not follow the 
coloring and AP rules of the majority perturb the evolution sufficiently to change the universality 
class.

\section{Discussion}

The main purpose of this paper was to explain in detail the reasons for the dramatic breakdown
of universality in agglomerative percolation on 2-d lattices. In finding this reason -- and 
demonstrating numerous other unexpected features of AP in these cases -- we indeed uncovered 
a new class of models with non-trivial symmetries. In the present paper only the simplest of 
these, having a $Z_2$ symmetry due to bipartivity, is treated in detail, while more complex 
situations leading to higher symmetries are only sketched. 

Agglomerative percolation is a very natural extension of the standard percolation model, and 
we expect a number of applications (some of which were already mentioned in \cite{Christen-2010}).
The main effect of bipartivity in AP is that the merging of large clusters is delayed, as compared 
to OP. It shares this feature with explosive percolation \cite{Achli-2009}, but in contrast 
to the latter this delay is not imposed artificially, but is a natural consequence of the 
structure of the model. Also, the merging of large clusters is not delayed in all circumstances,
but only subject to the symmetry structure imposed by bipartivity. The latter implies that 
clusters can have ``colors" (with $k$ colors in case of a $k-$partite network), and 
only the merging of clusters with different colors is delayed.

The effect of bipartivity is dramatic in case of 2-d lattices -- shifting, in particular,
the percolation threshold on infinite systems to the limit where the average cluster size
diverges and the number of clusters per site is zero. It is much less dramatic for 3-d lattices
(where we studied only the simple cubic lattice) and for random networks. In these cases
we see a clear effect, and the simulations indicate that universality with OP is broken,
but the percolation threshold is at finite values and the critical exponents are close to 
those of OP.

Future work is needed to settle these questions of universality. In particular, it would be 
of interest to study high-dimensional ($3 < d \leq 6$) simple hypercubic lattices, in order to 
see how the lattice models cross over to the random graph model. Another interesting subject 
for future work is a modified AP model (discussed briefly in Sec.~\ref{k-partite}) on the 
triangular lattice, where the relevant group is $Z_3$ instead of $Z_2$. Finally, there should 
exist a rich mathematical structure for modified AP models on $k-$partite networks with 
$k>3$, all of which is not yet understood.

\section*{Acknowledgements}

We thank Golnoosh Bizhani for numerous discussions, for helping with the simulations of 
AP on random bipartite graphs, and for carefully reading the manuscript.
\bibliography{mm}

\end{document}